\documentclass[a4paper,11pt, twoside]{article}
\usepackage[T1]{fontenc}
\usepackage[math]{iwona}

\pdfoutput=1
\usepackage[usenames,dvipsnames]{color}
\usepackage{hyperref}
\usepackage{longtable,epsfig,verbatim,multirow,multicol,hhline}
\usepackage{soul}
\usepackage{epstopdf,pslatex, pdfpages}
\usepackage{graphicx}
\usepackage[scaled]{helvet}
\usepackage[small]{caption}
\usepackage{subfigure}
\usepackage{eurosym}
\usepackage{sectsty}
\usepackage{natbib}
\usepackage{xspace}
\usepackage[left=2cm,right=2cm,top=2cm,bottom=2cm]{geometry}
\usepackage{fancyhdr}
\extrafloats{200}
\usepackage{longtable}
\usepackage[table]{xcolor}
\usepackage{floatrow}
\usepackage{enumitem}

\definecolor{FireBrick}{rgb}{0.70,0.13,0.13}

\makeatletter

\newcommand{\Msun}{\ensuremath{\mbox{M}_\odot}\xspace}

\newcommand{\gaia}{\textit{Gaia}\xspace}
\newcommand{\gaiap}{\textit{Gaia}'s~}

\newcommand{\theia}{\textit{Theia}\xspace}
\newcommand{\theiax}{\textit{Theia}}

\newcommand{\hips}{\textit{Hipparcos~}}
\newcommand{\hst}{\textit{HST}}
\newcommand{\NEATx}{\textit{NEAT}}

\newcommand{\plato}{\textit {PLATO} } 
\newcommand{\euclidx}{\textit {Euclid}} 
\newcommand{\euclid}{\textit {Euclid} } 
 
\newcommand{\jwst}{\textit{JWST}} 
\newcommand{\wfirst}{\textit{WFIRST}}
\newcommand{\hn}{H_0}

\newcommand{\remove}[1]{}

\DeclareCaptionFont{NavyBlue}{\color{NavyBlue}}
\DeclareCaptionFont{Blue}{\color{Blue}}

\urldef{\theiaurl}\url{http://theia.osug.fr}

\newcommand{\HPAversion}{version 20190805}

\begin{document}
\thispagestyle{empty}
\setcounter{page}{0}  
\begin{center}
  \textbf{\Large White paper for the Voyage 2050 long-term plan\\ in the ESA Science Programme}
  \vfill
  \fbox{%
    \begin{minipage}{0.9\linewidth}
      \begin{center}
        ~\\*[2em]
        \huge \textbf{Faint objects in motion:\\ the new frontier of high
          precision astrometry}\\*[2em]
        \large \textbf{Contact Scientist}: Fabien Malbet\\
        Institut de Plan\'etologie et d'Astrophysique de Grenoble (IPAG)\\
        \normalsize Universit\'e Grenoble Alpes, CS 40700,
        F-38058 Grenoble cedex 9, France \\
        \textit{Email:} \texttt{Fabien.Malbet@univ-grenoble-alpes.fr},
        \textit{Phone:} +33 476 63 58 33\\*[2em]
        ~
      \end{center}
    \end{minipage}}
  \vfill
  \vspace*{3em}
  \HPAversion
\end{center}
  
\clearpage
\twocolumn
\raggedbottom

\begin{center}
\fbox{\begin{minipage}{0.9\linewidth}
\Large \bf Faint objects in motion: the new frontier of high precision astrometry
\end{minipage}}
\end{center}

Sky survey telescopes and powerful targeted telesco\-pes play complementary roles in astronomy. In order \textbf{to investigate the nature and characteristics of the motions of very faint objects}, a flexibly-pointed instrument capable of high astrometric accuracy is an ideal complement to current astrometric surveys and a unique tool for precision astrophysics. \textbf{Such a space-based mission will push the frontier of precision astrometry from evidence of earth-massed habitable worlds around the nearest starts, and also into distant Milky way objects up to the Local Group of galaxies}. As we enter the era of the James Webb Space Telescope and the new ground-based, adaptive-optics-enabled giant telescopes, by obtaining these high precision measurements on key objects that \gaia could not reach, a mission that focuses on\textbf{ high precision astrometry science can consolidate our theoretical understanding of the local universe, enable extrapolation of physical processes to remote redshifts, and derive a much more consistent picture of cosmological evolution and the likely fate of our cosmos}.

Already several missions have been proposed to address the science case of faint objects in motion using high precision astrometry {ESA missions}: NEAT for M3, micro-NEAT for S1 mission, and \theia for M4 and M5 \citep{Boehm2017}.  Additional new mission configurations adapted with technological innovations could be envisioned to pursue accurate measurements of these extremely small motions. The goal of this white paper is to address the fundamental science questions that are at stake when we focus on the motions of faint sky objects and to briefly review quickly instrumentation and mission profiles.

\emph{Nota Bene: most Figures in this White Paper refers to the \theia specifications \citep[see][for details]{Boehm2017} which target at astrometric end-of-mission precisions of 10\,$\mu$as for faint object of $R=20$\,mag and 0.15\,$\mu$as for bright object of $R=5$\,mag (see Table \ref{tab:tech.summary.science})}.

\section{Science questions}
\label{sec:science-questions}

Europe has always been a pioneer of astrometry, from the time of ancient Greece to Tycho Brahe, Johannes Kepler, the Copernican revolution and Friedrich Bessel. ESA's \hips and \gaia satellites continued this tradition, revolutionizing our view of the Solar Neighborhood and Milky Way, and providing a crucial foundation for many disciplines of astronomy. \textbf{An unprecedented microarcsecond relative precision mission} will advance European astrometry still further, setting the stage for breakthroughs on \textbf{the most critical questions of cosmology, astronomy and particle physics}.

\subsection{Dark matter}
\label{sec:dark-matter}

\textbf{The current hypothesis of cold dark matter (CDM) urgently needs verification.} Dark matter (DM) is essential to the $\Lambda$ + CDM cosmological model ($\Lambda$CDM), which successfully describes the large-scale distribution of galaxies and the angular fluctuations of the Cosmic Microwave Background, as confirmed by the ESA / Planck mission. Dark matter is the dominant form of matter ($\sim 85\%$) in the Universe, and ensures the formation and stability of enmeshed galaxies and clusters of galaxies. The current paradigm is that dark matter is made of heavy, hence cold, particles; otherwise galaxies will not form. \textbf{However, the nature of dark matter is still unknown}.

There are a number of open issues regarding $\Lambda$CDM on small-scales. Simulations based on DM-only predict a 1) large number of small objects orbiting the Milky Way, 2) a steep DM distribution in their centre and 3) a prolate Milky Way halo. However, hydrodynamical simulations, which include dissipative gas and violent astrophysical phenomena (such as supernovae explosions and jets from galactic nuclei) can change this picture. Quantitative predictions are based on very poorly understood sub-grid physics and there is no consensus yet on the results. Answers are buried at small-scales, which are extremely difficult to probe. A new high precision astrometric mission appears to be the best way to settle the nature of DM and will allow us to validate or refute key predictions of $\Lambda$CDM, such as
\begin{itemize}[noitemsep,label=--]
\item the DM distribution in dwarf spheroidal galaxies  
\item the outer shape of the Milky Way DM halo 
\item the lowest masses of the Milky Way satellites and subhalos
\item the power spectrum of density perturbations
\end{itemize}
These observations will significantly advance research in\-to DM. They may indicate that DM is warmer than $\Lambda$CDM predicts. Or we may find that DM is prone to self-interac\-tions that reduces its density in the central part of the satellites of the Milky Way. We may discover that DM has small interactions that reduce the number of satellite companions. Alternatively, measurement of the Milky Way DM halo could reveal that DM is a sophisticated manifestation of a modification of Einstein's gravity.

\subsubsection{ The DM distribution in dwarf spheroidal galaxies}

\begin{figure}[t]
  \centering
  \includegraphics[width=0.8\textwidth]{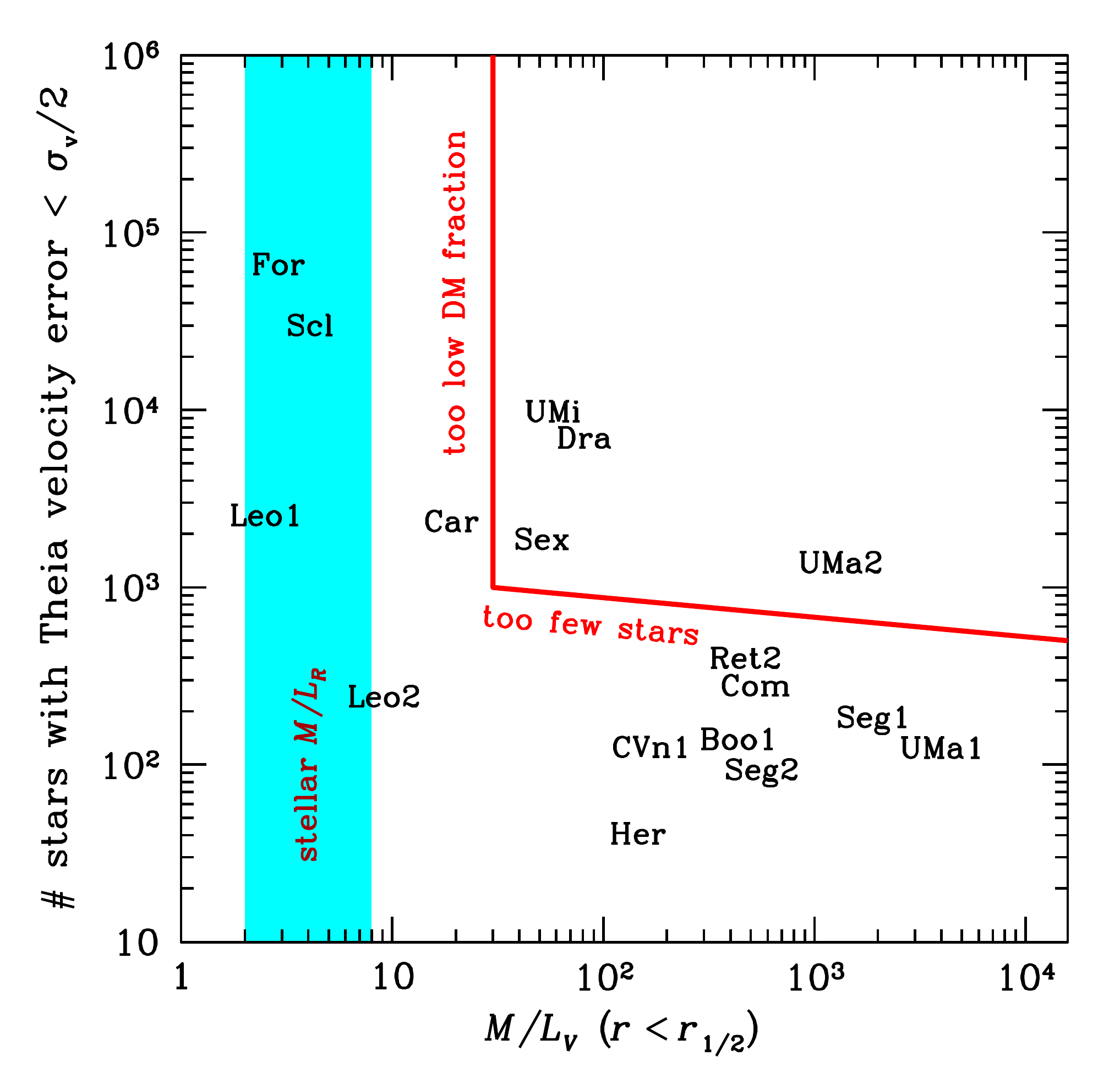}
  \caption{\em Number of dwarf spheroidal galaxy stars within a high precision astrometry missionfield with expected plane-of-sky errors lower than half the galaxy's velocity dispersion as a function of the galaxy's estimated mass-to-light ratio within the effective (half-projected-light) radius of the galaxy.  Luminosities and total masses within the half-light radii are mainly from \citet{Walker+09}.} 
  \label{fig:NvsMoverL}
\end{figure}

\begin{figure*}[t]
\centering
\includegraphics[width=0.45\textwidth,clip]{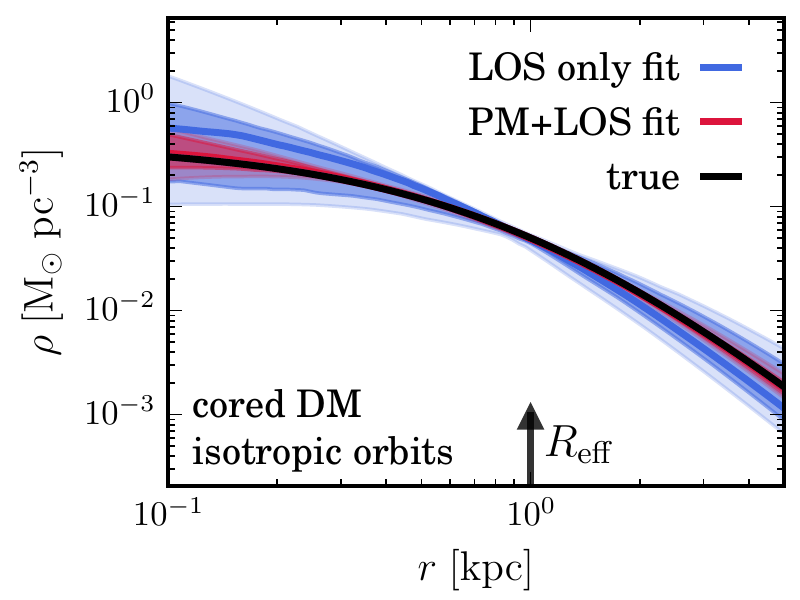}
\includegraphics[width=0.45\textwidth,clip]{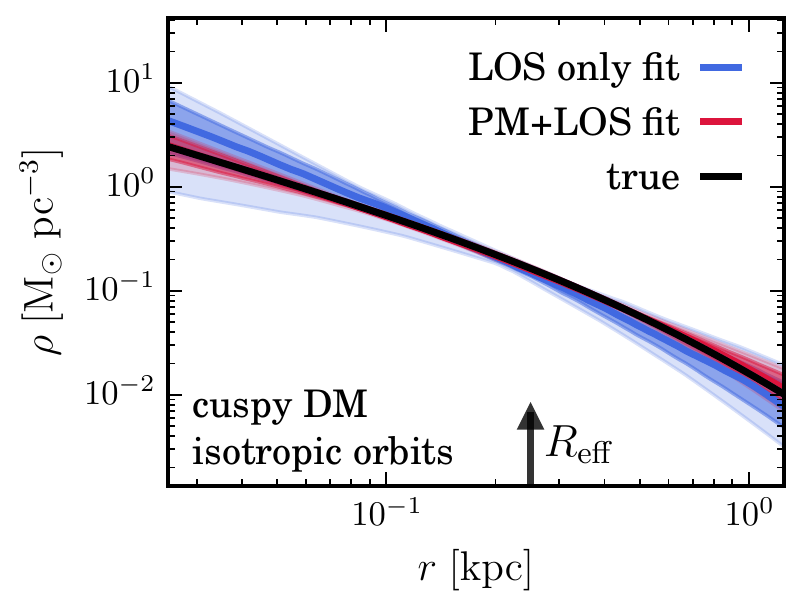}
\includegraphics[width=0.45\textwidth,clip]{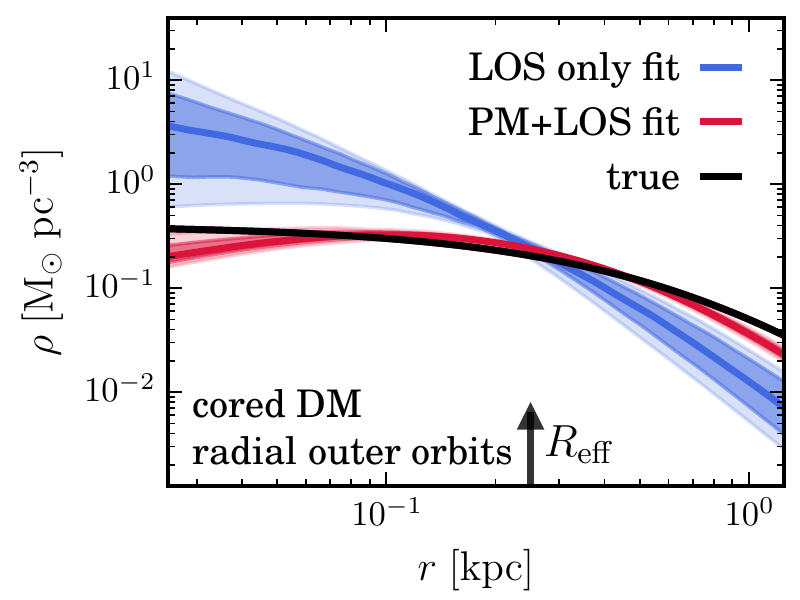}
\includegraphics[width=0.45\textwidth,clip]{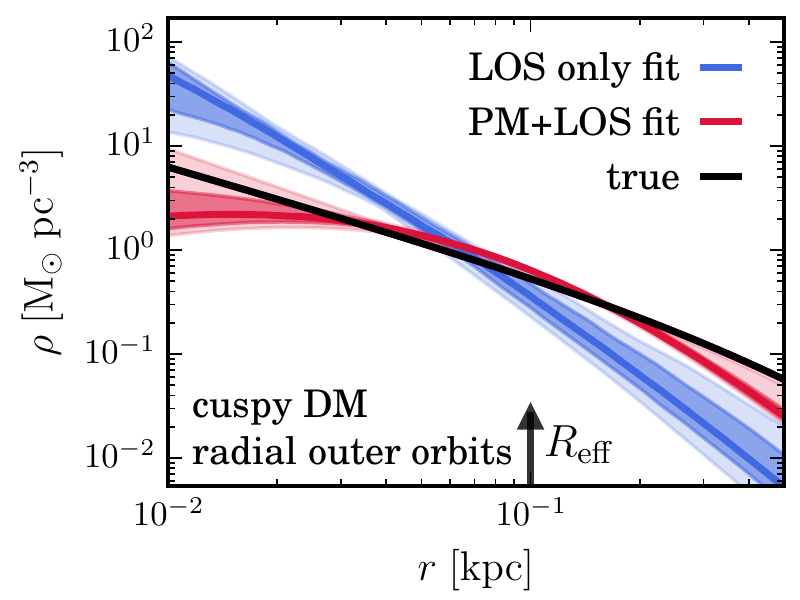}
\caption{\em Reconstruction of the DM halo profile of the Draco dSph without (\textit{blue}) and with (\textit{red}) proper motions using the mass-orbit modeling algorithm of Watkins et al. (2013). Four mocks of Draco were used, with cored (\textit{left}) and cuspy (\textit{right}) DM halos, and with isotropic velocities everywhere (\textit{top}) or only in the inner regions with increasingly radial motions in the outer regions (\textit{bottom}). The effective (half-projected light) radii of each mock is shown with the \textit{arrows}. The stellar proper motions in the mocks were  perturbed with apparent magnitude dependent errors as expected with 1000 hours of observations spread over 4 years.}
\label{fig:rhoofrPM}
\end{figure*}

Because they are DM-dominated (see Fig.~\ref{fig:NvsMoverL} where the number of stars versus the mass-to-light ratio is present\-ed), dwarf Spheroidal galaxies (dSphs) are excellent laboratories to test the distribution of DM within the central part of small galaxies and disentangle the influence of complex baryonic processes from that of DM at these scales.

Simulations from \citet{Onorbe+15} or \citet{Read+16} for example show that the DM distribution (referred to as DM profile) in dSphs strongly depends on their star formation history. More specifically, these simulations find that CDM can be heated by bursty star formation inside the stellar half light radius $R_{1/2}$, if star formation proceeds for long enough. As a result, some dSphs like Fornax have formed stars for almost a Hubble time and so should have large central DM cores, while others, like Draco and Ursa Major2, had their star formation truncated after just $\sim 1-2$\,Gyrs and should retain their steep central DM cusp.

Large DM cores could also be  attributed however to strong self-interactions. Hence finding evidence for such cores in the faintest dSphs (which are even more DM dominated \citep{Wolf+10} than the classical ones), will bring tremendous insights about the history of baryonic process\-es in these objects and could even dramatically change our understanding of the nature of DM. Indeed, self-inter\-acting DM \citep{Spergel&Steinhardt00} is expected to scatter in the dense inner regions of dSphs, and thus leads to homogeneous cores. Finding such a core DM distribution in dSphs could then reveal a new type of particle forces in the DM sector and provide us with new directions to  build extensions of Standard Model of particle physics. On the other hand, finding cuspy DM profiles in all dSphs (including the faintest ones) will confirm $\Lambda$CDM and place strong constraints on galaxy formation. As shown in Figs.~\ref{fig:vposerrvsDist} and \ref{fig:astroComparisons}, a telescope with micro-arcsecond astrometric precision allows us to determine whether the DM distribution in dSphs is cuspy or has a core, and hence can lead to a very significant breakthrough regarding the nature of DM.
 
To determine the inner DM distribution in dSphs, one needs to remove the degeneracy between the radial DM profile and orbital anisotropy that quantifies whether stellar orbits are more radial or more tangential in the Jeans equation \citep{Binney&Mamon82}.  This can be done by adding the proper motions of stars in dSphs.  Fig.~\ref{fig:rhoofrPM} shows that for the Draco dSph (which was obtained using single-component spherical mock datasets from the \gaia Challenge Spherical and Triaxial Systems working group,\footnote{See \url{http://astrowiki.ph.surrey.ac.uk/dokuwiki/doku.php?id=tests:sphtri}} and the number of stars expected to be observed by a high precision astrometry mission), the inclusion of proper motions lifts the cusp / core degeneracy that line-of-sight-only kinematics cannot disentangle.

We remark in addition that a high precision astrometric mission is able to perform follow-ups of \gaiap observations of dSphs streams of stars if needed. Not only will such a mission provide the missing tangential velocities for stars with existing radial velocities, but it will also provide crucial membership information - and tangential velocities - for stars in the outer regions of the satellite galaxies that are tidally disrupted by the Milky Way.

\subsubsection{ The triaxiality of the Milky Way dark matter halo \label{triaxial} }

For almost two decades cosmological simulations have shown that Milky Way-like DM halos have triaxial shapes, with the degree of triaxiality varying with radius \citep[][for example]{dubins_94, kkzanm04}: halos are more round or oblate at the center, become triaxial at intermediate radii, and prolate at large radii \citep{zemp_etal_11}. 

Precise measurement of the velocity of distant Hyper Velocity Stars (hereafter HVS) can test these departures from spherical symmetry, independently of any other technique attempted so far (such as the tidal streams). HVSs were first discovered  serendipitously \citep{Brown+05,Hirsch+05,Edelmann+05}, and later discovered in a targeted survey of blue main-sequence stars \citep[][and references therein]{Brown15_ARAA}. \gaia  measurements demonstrate that candidate HVSs include unbound disk runaways \citep{Irrgang2019}, unbound white dwarfs ejected from double-degenerate type Ia supernovae \citep{Shen2018}, and runaways from the LMC \citep{Erkal2018}, however the highest-velocity main sequence stars in the Milky Way halo have trajectories that point from the Galactic center \citep{Brown2018, Koposov2019}.
\vspace{5mm}

\begin{figure}[thb]
\centering
\fbox{\includegraphics[width=0.9\textwidth,clip]{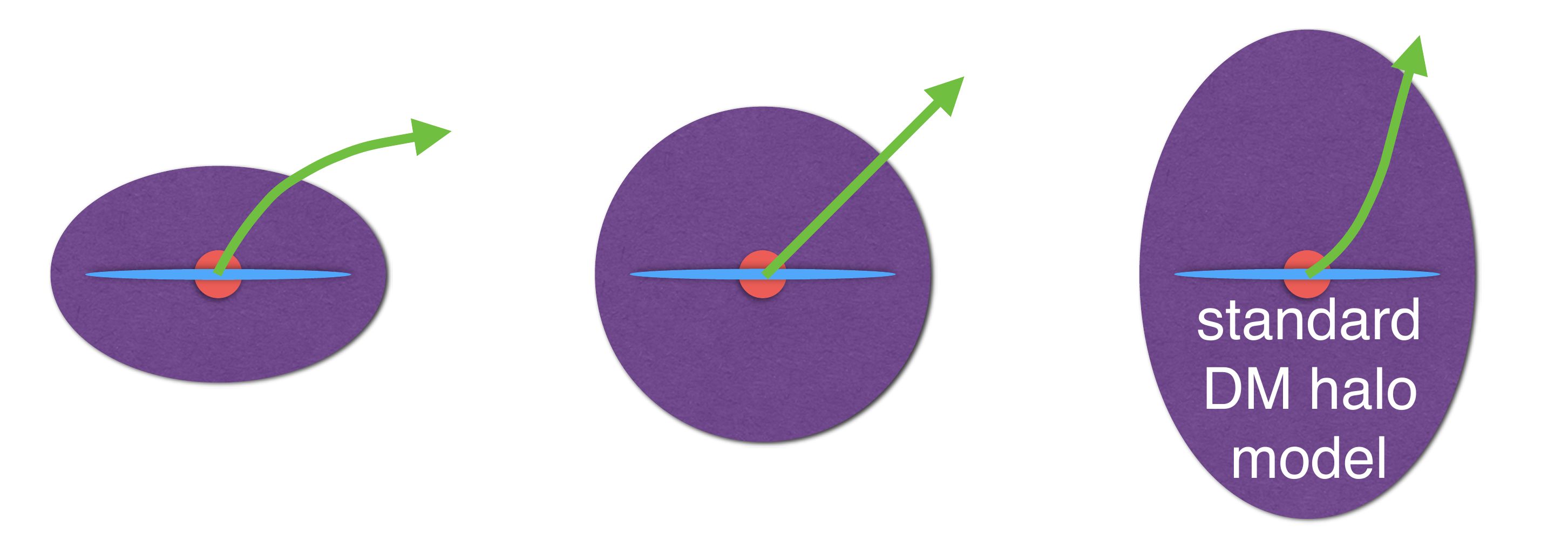}}
\caption{\em Illustration of the trajectories of hyper velocity stars ejected from Galactic Centre for 3 different outer
 DM halo shapes: oblate (\textit{left}), spherical (\textit{middle}), and prolate (\textit{right}).} 
\label{Fig:triaxial_halos}
\end{figure}

Because these velocities exceed the plausible limit for a runaway star ejected from a binary, in which one component has undergone a supernova explosion, the primary mechanism for a star to obtain such an extreme velocity is assumed to be a three-body interaction and ejection from the deep potential well of the supermassive black hole at the Galactic center \citep{Hills88,Yu&Tremaine03}.

\vspace{8mm}
\begin{figure}[bht]
    \includegraphics[width=0.8\textwidth,trim=0cm 0cm 0cm 1.5cm,]{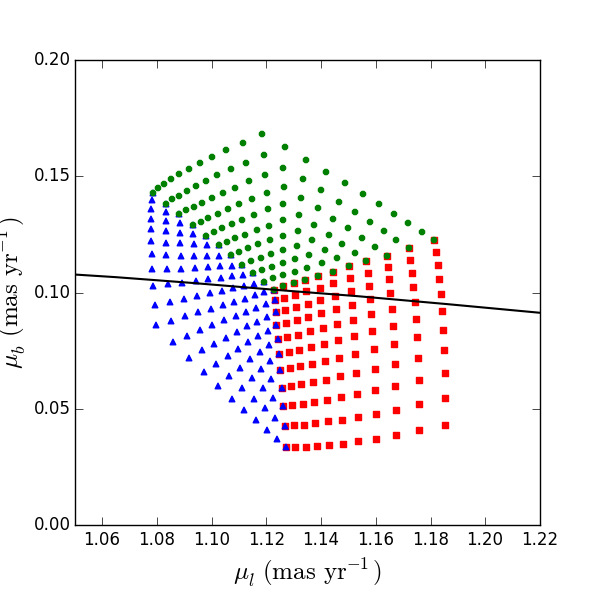}
\caption{\em Expected proper motions of HVS5 under different assumptions about the shape and orientation of the DM halo. The families of models are shown with the halo major axis along the Galactic X- (\textit{red squares}), Y- (\textit{blue triangles}), and Z-   (\textit{green circles}) coordinates.  The \textit{solid line} shows how the centroid of the proper motions will shift with a different distance to HVS5.}
\label{fig:pm_hvs5} 
\end{figure}

By measuring the three-dimensional velocity of these stars, we will reconstruct the triaxiality of the Galactic potential. In a spherical potential, unbound HVS ejected from the Galactic center should travel in nearly a straight line, as depicted in Fig.\ref{Fig:triaxial_halos}.  However, for triaxial halos, the present velocity vector should not point exactly from the Galactic Center because of the small curvature of the orbit caused by non-spherically symmetric part of the potential \citep{Gnedin+05,Yu&madau07}. While both the halo and stellar disc induce transverse motions, the effect is dominated by halo triaxiality at the typical distance of HVS. The deflection contributed by the disc peaks around 10 kpc but quickly declines at larger distances, while the deflection due to the triaxial halo continues to accumulate along the whole trajectory. Fig.~\ref{fig:pm_hvs5} actually shows the spread of proper motion for one star, HVS5, for different halo shapes (different halo axis ratios and different orientations of the major axis).

Proper motions of several HVSs were measured with the Hubble Space Telescope (\textit{HST}) by \cite{Brown+15}, using an astrometric frame based on background galaxies. However, these measurements were not sufficiently accurate to constrain the halo shape or the origin of HVS.  A high precision astrometric mission with a sufficiently large field of view could include about 10 known quasars from the SDSS catalog around most HVSs. This will provide a much more stable and accurate astrometric frame, and will allow us to constrain the halo axis ratios to about 5$\%$.

\begin{figure}[bht]
\centering
\includegraphics[width=0.9\textwidth, trim=0.cm 0.3cm 0.8cm 1cm ,clip]{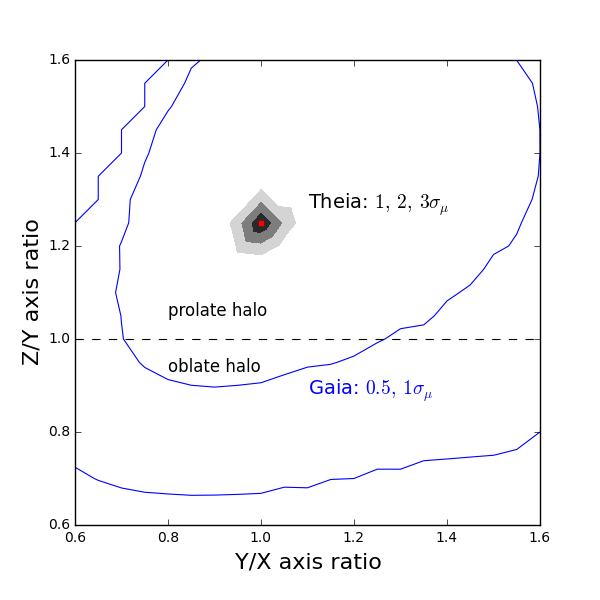}
\caption{\em Example of a reconstruction of the Galactic halo shape from  a high precision astrometry mission measurement of proper motion of HVS5.  The assumed proper motions correspond to a prolate model with $q_X = q_Y = 0.8\, q_Z$, marked by a \textit{red square}. \textit{Shaded contours} represent confidence limits corresponding to the expected 1, 2, and $3\, \sigma_\mu$ proper motion errors. The \textit{outer blue contours} show the accuracy that will be achieved by \gaia at the end of its mission, even if its expected error was reduced by a factor of 2.} \label{Fig:axishvs5d20theia} 
\end{figure}

\begin{figure*}[t]
\centering
\includegraphics[width=0.28 \textwidth]{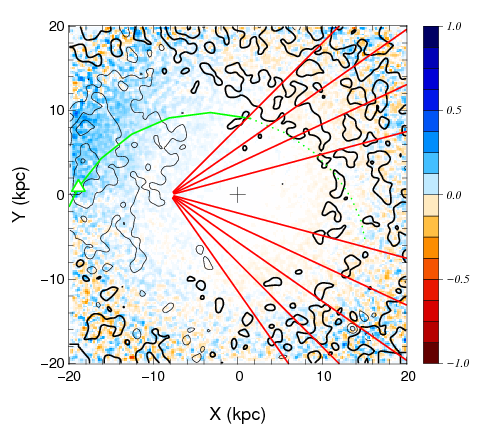}
\includegraphics[width=0.28 \textwidth]{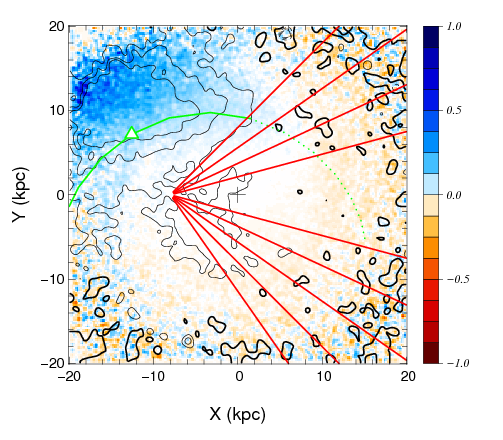}
\includegraphics[width=0.28 \textwidth]{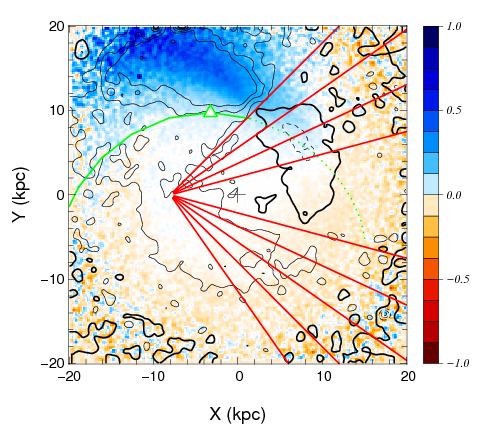}
\includegraphics[width=0.28 \textwidth]{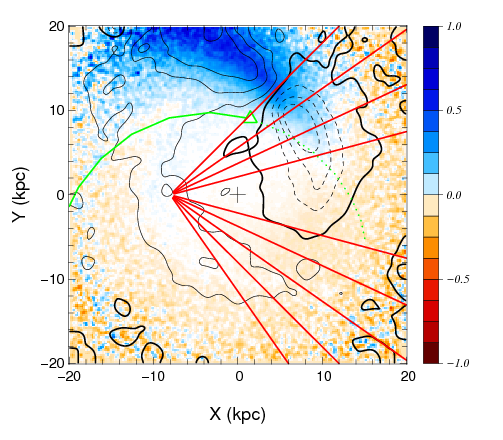}
\includegraphics[width=0.28 \textwidth]{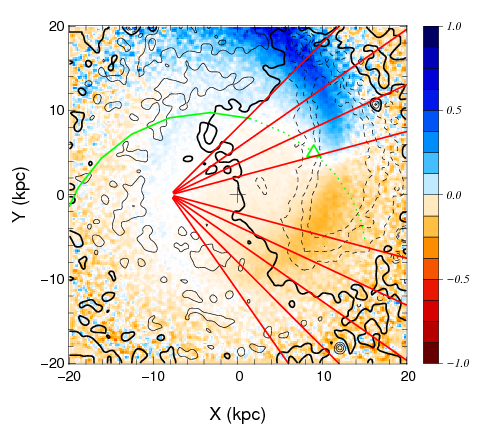}
\includegraphics[width=0.28 \textwidth]{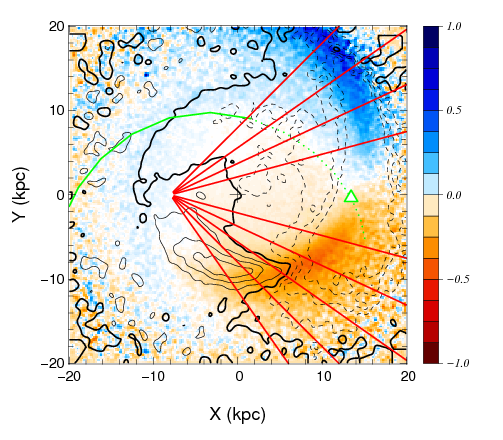}
\caption{\em Face-on view of the evolution of the perturbation of a Galactic Disc due to a DM subhalo of mass 3$\%$ of the mass of the disc crossing the disc from above. The \textit{upper} and \textit{lower} panels are before and after the crossing, respectively, for different times 125, 75 and 25 Myr before the crossing and 25,75,125 Myr after (from \textit{left} to \textit{right}).  The mean displacement amplitude is indicated in the color bar, while the \textit{contours} indicate the amplitude of the bending mode in velocity space,  using plain lines for positive values and dashed lines for negative values. The \textit{green line} shows the projected orbit of the subhalo (dashed line after the impact with the disc). The \textit{green triangle} shows the current location of the subhalo on its orbit.  The \textit{red lines} are our potential lines of sight for \theia, spaced by 10$^\circ$ in longitude with one pointing above the plane and one below the plane, that will allow us to map the disc perturbation behind the Galactic Center. }
\label{fig:dm:1}
\end{figure*}

Fig.\ref{Fig:axishvs5d20theia} shows indeed that with a precision of $4 \, \mu$as/yr one can constrain the orientation of the halo major axis and measure the axis ratios to an accuracy of $\delta (q_Z/q_X) < 0.05$ for the typical HVS distance of 50 kpc.  For comparison, \gaia at the end of its mission will achieve only $40 - 150 \, \mu$as/yr, which is highly insufficient to provide useful constraints on the axis ratios.

Statistical studies of high-precision proper motions of HVSs can also constrain departures of the halo shape from spherical (Gallo, Ostorero,\& Diaferio, in preparation). Indeed, numerical simulations of the trajectories of synthetic HVSs ejected through the Hills mechanism show that the distributions of the HVS tangential velocities in the Galactocentric reference frame are significantly different from spherical and non-spherical halos: the significance is $P \le 1.3e-6 $ for oblate halos with $q_Z/q_X \le 0.9$ and $P \le 2.2e-5$ for prolate halos with $q_Z/q_X \ge 1.1$ . The median tangential velocity of a sample of $\sim 100$ HVSs located at heliocentric distances $\sim 50$ kpc can differ by $\sim 5-10 km/s$, implying differences in proper motions of $\sim 20-40$ $\mu$as/yr between spherical and non-spherical halos.

Finally, an accurate measurement of HVS velocities may lead to improved understanding of the black hole(s) at the Galactic center. Indeed, theoretical models show that HVSs will have a different spectrum of ejection velocities from a binary black hole versus a single massive black hole. \gaia has led to the discovery of several hypervelocity stars \citep[ejection velocities of over 550 km/s][]{Irrgang2018,Hattori2019, Irrgang2019}, that were definitely not ejected from the Galactic Center but were ejected from spiral arms in the MW disk. These most likely point to intermediate mass black holes of mass $~$100 Msun - these could be local remnants of binary BH mergers of the kind discovered by LIGO and could have important implications for our understanding of stellar evolution.

\subsubsection{Orbital distribution of Dark Matter from the orbits of halo stars}

The orbits of DM particles in halos\footnote{For an analysis of orbital content of DM halos see \cite{valluri_etal_10, valluri_etal_12,bryan_etal_12,valluri_etal_13}.} cannot be detected directly since DM particles interact only weakly with normal matter. However, in a triaxial potential such as described above, it is expected that a large fraction of the DM orbits do have any net angular momentum. Hence these particles should get arbitrarily close to the center of the cusp, regardless of how far from the center they were originally. This allows DM particles, which annihilate within the cusp to be replenished on a timescale $10^4$ longer than in a spherical halo \citep[analogous to loss cone filling in the case of binary black holes][]{merritt_poon_04}.

Recent work on the orbital properties and kinematic distributions of halo stars and DM particles show that halo stars, especially the ones with lowest metallicities, are relatively good tracers of DM particles \citep{2013ApJ...767...93V, 2018PhRvL.120d1102H, 2018JCAP...04..052H} and observations with \gaia DR2 may have already led to the kinematic discovery of dark substructure \citep{Necib2019}.
The orbits reflect both the accretion/formation history and the current shape of the potential because DM halos are dynamically young (i.e. they are still growing and have not attained a long term equilibrium configuration where all orbits are fully phase mixed).  This opens up the very exciting possibility that one can infer the kinematical distribution of DM particles by assuming that they are represented by the kinematics of halo stars.

\subsubsection{Perturbations by Dark Matter subhalos}

A central prediction of $\Lambda$CDM in contrast to many alternatives of DM, such as warm DM \citep[e.g.][]{Schaeffer:1984bt} or interacting DM \citep[e.g.][]{Boehm:2014vja}, is the existence of numerous $10^6$ to $10^8$ M$_\odot$ DM \emph{subhalos} in the Milky Way halo. Their detection is extremely challenging, as they are very faint and lighter than dSphs. However, N-body simulations of the Galactic Disc show that such a DM halo  passing through the Milky Way disc will warp the disc and produce a motion (bending mode), as shown in Fig.~\ref{fig:dm:1}. This opens new avenues for detection as such perturbations of the disc will result in anomalous motions of the stars in the disc \citep[e.g.][for recent analysis]{2015MNRAS.446.1000F}, that could give rise to an astrometric signal.  

\begin{figure*}[ht]
\centering
\includegraphics[width=0.4\textwidth, clip]{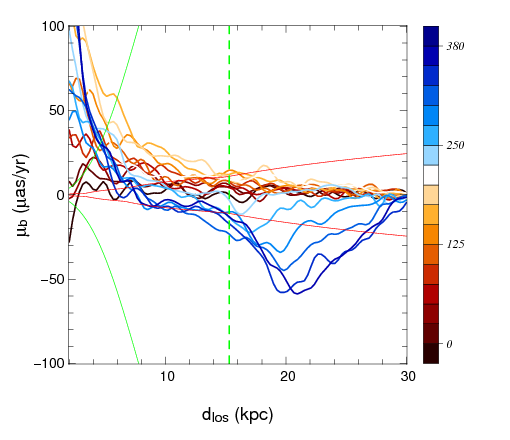}
\includegraphics[width=0.4 \textwidth,clip]{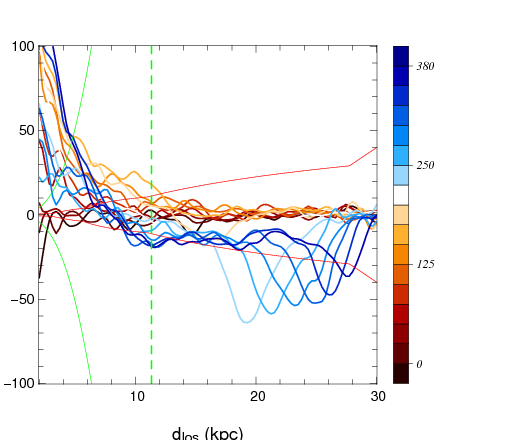}
\caption{\em Astrometric signatures in the proper motion along Galactic latitude of the perturbation of disc stars by a subhalo. The \textit{left} and \textit{right} panels show lines of sight as a function of distance along the line of sight and time,  for $\ell=-25^\circ$  and $\ell=+25^\circ$ 
respectively for \remove{a Galactic latitude of} $b=+2^\circ$. The color codes the time in Myr, \textit{red} for times prior to the crossing of the plane by the satellite, \textit{blue} for later times. The \textit{green line} is \gaia's expected end of mission performance for a population of red clump stars along these lines of sight. The \textit{vertical dashed line} is \gaia's detection limit ($G$=20) for the same population. The \textit{red lines} are \theia's expected 1$\sigma$ accuracy  for the same stars and for a 400~h exposure of the field over the course of the mission.
}
\label{fig:dm:2}
\end{figure*}

These anomalous bulk motions develop both in the solar vicinity \citep{2012ApJ...750L..41W} and on larger scales \citep{2015MNRAS.446.1000F}, see Fig.\ref{fig:dm:2}. Therefore, measuring very small proper motions of individual faint stars in different directions towards the Galactic disc could prove the existence of these subhalos and confirm the CDM scenario. Alternatively, in case they are not found, high precise astrometric observations will provide tantalizing evidence for alternative DM scenarios, the most popular today being a warmer form of DM particle, though these results could also indicate DM interactions \citep{Boehm:2014vja}.   

A field of view of $1^\circ\times 1^\circ$ in the direction of the Galactic disc has $\sim 10^6$ stars with an apparent magnitude of $R \leq 20$ (given by the confusion limit). Given the astrometric precisions per field of view of Fig.~\ref{fig:astroComparisons}, a high precision astrometric instrument could detect up to 7 impacts on the disc from sub-halos as small as a few $10^6\,\Msun$.

\gaia DR2  astrometry has led to the discovery of gaps in tidal streams \citep{2018ApJ...863L..20P} like the GD1 stream. The gaps and off-stream stars (spur) are consistent with gravitational interactions with compact DM subhalos. Further more, \gaia DR2 data has revealed that globular cluster streams (GD1 and Jhelum) show evidence for cocoon like structures that most likely arise from evolution inside a (dark) subhalo prior to their tidal disruption by the Milky Way itself \citep{Carlberg2018, Malhan2019, Bonaca2019}.  The high astrometric precision of a {\theia}-like mission will enable us to measure the small velocity perturbations around the gaps in streams and allow for a much more accurate determination of both the masses and density structures of the perturbing dark subhalos. 

\subsubsection{ Ultra-compact minihalos of dark matter in the Milky Way} 
\label{sec:science_minihalos}

In the $\Lambda$CDM model, galaxies and other large-scale structures formed from tiny fluctuations in the distribution of matter in the early Universe. Inflation predicts a spectrum of primordial fluctuations in the curvature of spacetime, which directly leads to the power spectrum of initial density fluctuations.  This spectrum is observed on large scales in the cosmic microwave background and the large scale structure of galaxies, but is very poorly constrained on scales smaller than 2\,Mpc.  This severely restricts our ability to probe the physics of the early Universe.  A high precision astrometric mission could provide a new window on these small scales by searching for astrometric microlensing events caused by \emph{ultra-compact minihalos} (UCMHs) of DM.

UCMHs form shortly after matter domination (at $z\sim1000$), in regions that are initially overdense \citep[e.g. $\delta\rho/\rho > 0.001$ in][]{2009ApJ...707..979R}.  UCMHs only form from fluctuations about a factor of 100 larger than their regular cosmological counterparts, so their discovery will indicate that the primordial power spectrum is not scale invariant.  This will rule out the single-field models of inflation that have dominated the theoretical landscape for the past thirty years.  Conversely, the absence of UCMHs can be used to establish upper bounds on the amplitude of the primordial power spectrum on small scales \citep{Bringmann11}, which will rule out inflationary models that predict enhanced small-scale structure \citep{Aslanyan16}.

\begin{figure}[htb]
  \includegraphics[width=0.8\textwidth]{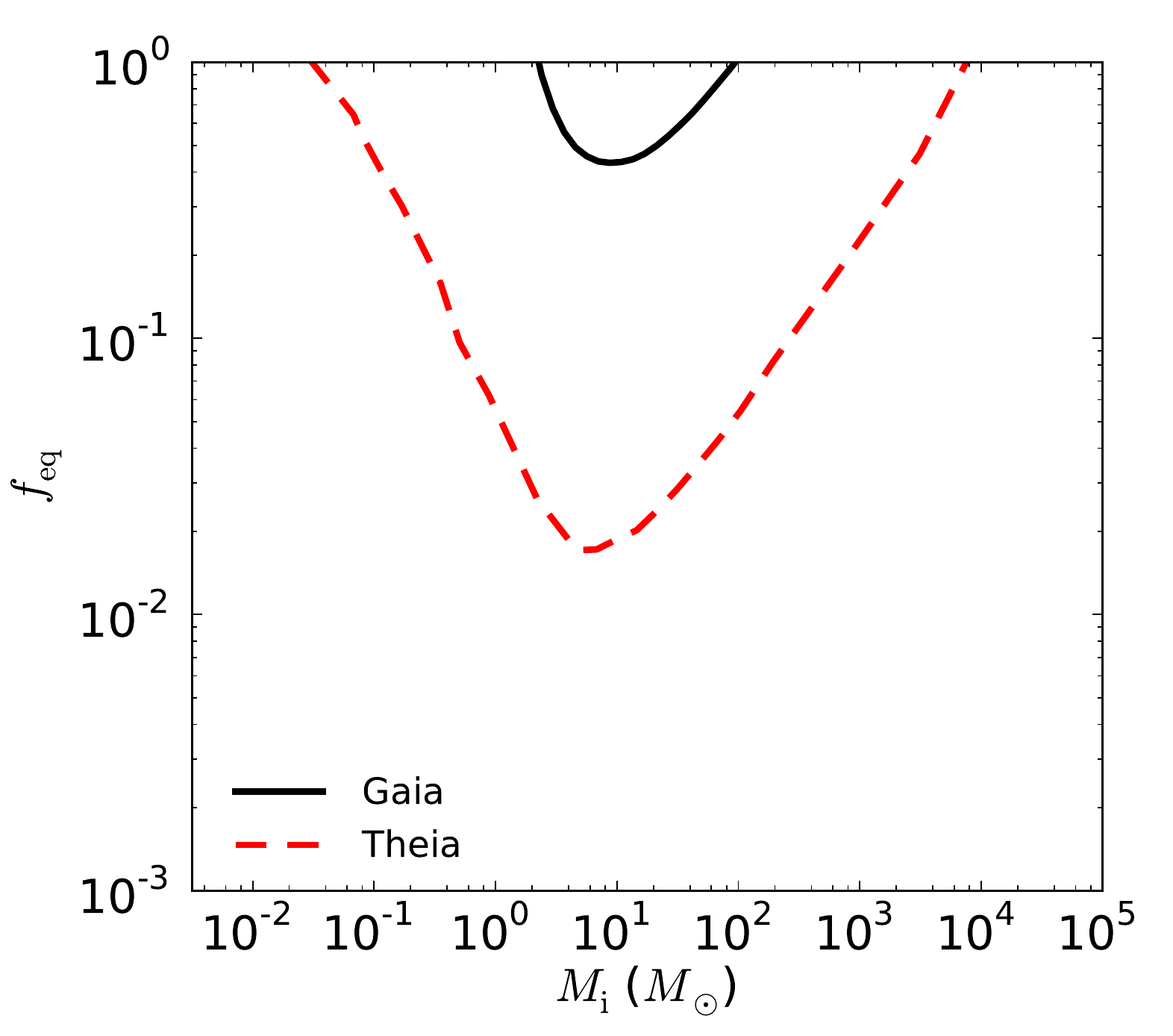}
  \caption{\em Projected sensitivity of  a high precision astrometry missionto the fraction of dark matter in the form of ultracompact minihalos (UCMHs) of mass $M_i$ at the time of matter-radiation equality.  Smaller masses probe smaller scales, which correspond to earlier formation times (and therefore to \textit{later} stages of inflation). A UCMH mass of 0.1\,M$_\odot$ corresponds to a scale of just 700\,pc.  Expected constraints from \gaia are given for comparison, showing that a {\theia}-like mission will provide much stronger sensitivity, as well as probe smaller scales and earlier formation times than ever reached before.}
  \label{fig:ucmh}
\end{figure}

\begin{figure*}[t]
  \includegraphics[width=0.8\textwidth,trim = 0cm 6.3cm 0cm 7.5cm,clip]{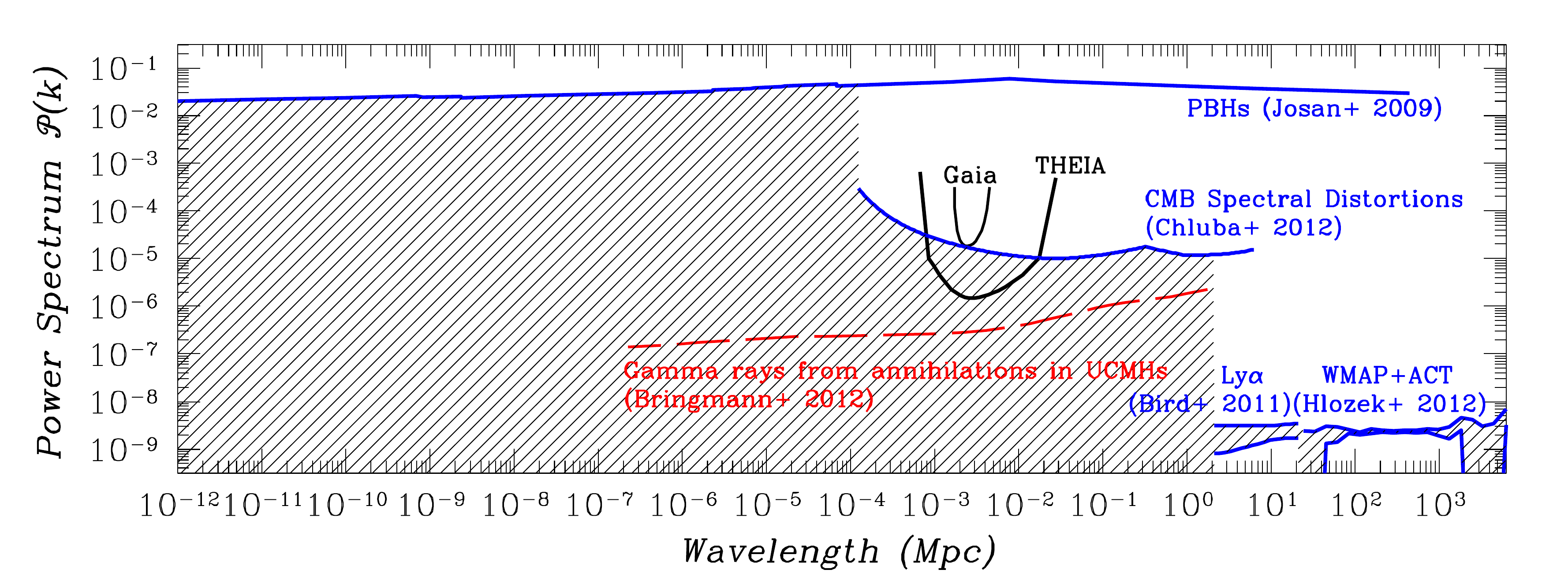}
  \caption{\em Limits on the power of primordial cosmological perturbations at all scales, from a range of different sources.  a {\theia}-like  mission will  provide far stronger sensitivity to primordial fluctuations on small scales than \gaia, spectral distortions or primordial black holes (PBHs).  Unlike gamma-ray UCMH limits,  a high precision astrometry mission's sensitivity to cosmological perturbations will also be independent of the specific particle nature of dark matter.}
  \label{fig:ucmh2}
\end{figure*}

Like standard DM halos, UCMHs are too diffuse to be detected by regular photometric microlensing searches for MAssive Compact Halo Objects (MACHOs).  Because they are far more compact than standard DM halos, they however produce much stronger \textit{astrometric} microlensing signatures \citep{2012PhRvD..86d3519L}.  By searching for microlensing events due to UCMHs in the Milky Way, a high precision astrometric mission will provide a new probe of inflation.  A search for astrometric signatures of UCMHs in the \gaia dataset could constrain the amplitude of the primordial power spectrum to be less than about $10^{-5}$ on scales around 2\,kpc \citep{2012PhRvD..86d3519L}.  Fig.\ \ref{fig:ucmh} shows that higher astrometric precision (corresponding to that of Fig.~\ref{fig:astroComparisons}) will provide more than \textit{an order of magnitude higher sensitivity} to UCMHs, and around \textit{four orders of magnitude greater mass coverage} than \gaia.  These projections are based on 8000\,hr of observations of 10 fields in the Milky Way disc, observed three times a year, assuming that the first year of data is reserved for calibrating stellar proper motions against which to look for lensing perturbations.  Fig.\ \ref{fig:ucmh2} shows that a high precision astrometric mission will test the primordial spectrum of perturbations down to scales as small as 700\,pc, and improve on the expected limits from \gaia by over an order of magnitude at larger scales.

The results will be independent of the DM nature, as astrometric microlensing depends on gravity only, unlike other constraints at similar scales based on DM annihilation, from the \textit{Fermi} Gamma Ray Space Telescope \citep{Bringmann11}.  An astrometric mission with higher precision (shown in  Fig.~\ref{fig:astroComparisons})  will have sensitivity  four orders of magnitude stronger than constraints from the absence of primordial black holes (PBHs), and more than an order of magnitude better than CMB spectral distortions \citep{2012ApJ...758...76C}, which give the current best model-independent limit on the primordial power spectrum at similar scales.

\nocite{2011MNRAS.413.1717B, 2012ApJ...749...90H}

\subsubsection{Directly Testing Gravity} 

Using the nearest star, Proximan Cen, astrometry could measure the behaviour of gravity at low accelerations.  A high precision astrometry mission with an extended baseline of 10 years and a precision of 0.5 $\mu$as could measure the wide binary orbit of Proxima Centauri around alpha Centauri A and B to distinguish between Newtonian gravity and Milgromian dynamics (MOND).  The separation between Proxima cen and the Alpha Centauri system suggests orbital acceleration that is significantly less than MOND acceleration constant $a_0 \sim 1.2 \times 10^{-10}$ m/s$^2$ \citep{Banik2019}.  It would be the first direct measurement of the departure from Newtonian gravity in the very weak field limit, as expected in MOND, and the results could have profound implications on fundamental physics.

\subsection{Exoplanets}
\label{sec:exoplanets}

\subsubsection{The Frontier of Exoplanet Astrophysics}

The ultimate exoplanetary science goal is to answer the enigmatic and ancient question, ``\emph{Are we alone?'}' via unambiguous detection of biogenic gases and molecules in the atmosphere of an Earth twin around a Sun-like star \citep{Schwieterman+16}. Directly addressing the age-old questions related to the uniqueness of the Earth as a habitat for complex biology constitutes today the vanguard of the field, and it is clearly recognized as one unprecedented, cross-technique, interdisciplinary endeavor.

Since the discovery of the first Jupiter-mass companion to a solar-type star \citep{Mayor&Queloz95}, tremendous progress has been made in the field of exoplanets. Our knowledge is expanding ever so quickly due to the discovery of thousands of planets, and the skillful combination of high-sensitivity space-borne and ground-based programs that have unveiled the variety of planetary systems architectures that exist in the Galaxy \citep[e.g.][]{2013Sci...340..572H,2011arXiv1109.2497M}.  Preliminary estimates \citep[e.g.][]{Winn&Fabrycky15} are now also available for the occurrence rate $\eta_\uplus$ of terrestrial-type planets in the Habitable Zone (HZ) of stars more like the Sun ($\eta_\uplus\sim10\%$) and low-mass M dwarfs ($\eta_\uplus\sim50\%$).

However, transiting or Doppler-detected HZ terrestrial planet candidates (including the recent discovery of the $m_{\rm p}\sin i=1.3$ $M_\oplus$ HZ-planet orbiting Proxima Centauri) lack determinations of their bulk densities $\varrho_{\rm p}$. Thus, the HZ terrestrial planets known to-date are not amenable to make clear statements on their habitability.  The \textit{K2}, \textit{TESS}, and \textit{PLATO} missions are bound to provide tens of Earths and Super Earths in the HZ around bright M dwarfs and solar-type stars for which $\varrho_{\rm p}$ estimates might be obtained in principle, but atmospheric characterization for the latter sample might be beyond the capabilities of JWST and the Extremely Large Telescopes (ELTs). The nearest stars to the Sun are thus the most natural reservoir for the identification of potentially habitable rocky planets that might be characterized via a combination of high-dispersion spectroscopy and high-contrast imaging with the ELTs \citep{Snellen+15} or via coronagraphic or interferometric observations in space \citep{Leger15}.

 Unlike the Doppler and transit methods, astrometry alone can determine reliably and precisely the true mass and  three-dimensional orbital geometry of an exoplanet, which are fundamental inputs to models of planetary evolution, biosignature identification, and habitability. 
By determining the times, angular separation and position angle at periastron and apoastron passage,
exquisitely precise astrometric position measurements will allow the prediction of where and when a planet will be at its brightest (and even the likelihood of a transit event), thus (a) crucially helping in the optimization of direct imaging observations and (b) relaxing important model degeneracies in predictions of the planetary phase function in terms of orbit geometry, companion mass, system age, orbital phase, cloud cover, scattering mechanisms and degree of polarization \citep[e.g.][]{Madhusudhan&Burrows12}.  \emph{Only a high precision astrometric mission's observations  will have the potential to 1) discover most of the potentially habitable planets around the nearest stars to the Sun, 2) directly measure their masses and system architectures, and 3) provide the most complete target list and vastly improve the efficiency of detection of potential habitats for complex exo-life with the next generation of space telescopes and ELTs.}

\subsubsection{Fundamental Program} 

\begin{figure*}[tb]
\centering
\includegraphics[width=0.65 \textwidth, trim = 0cm 0cm 0cm 1cm, clip]
{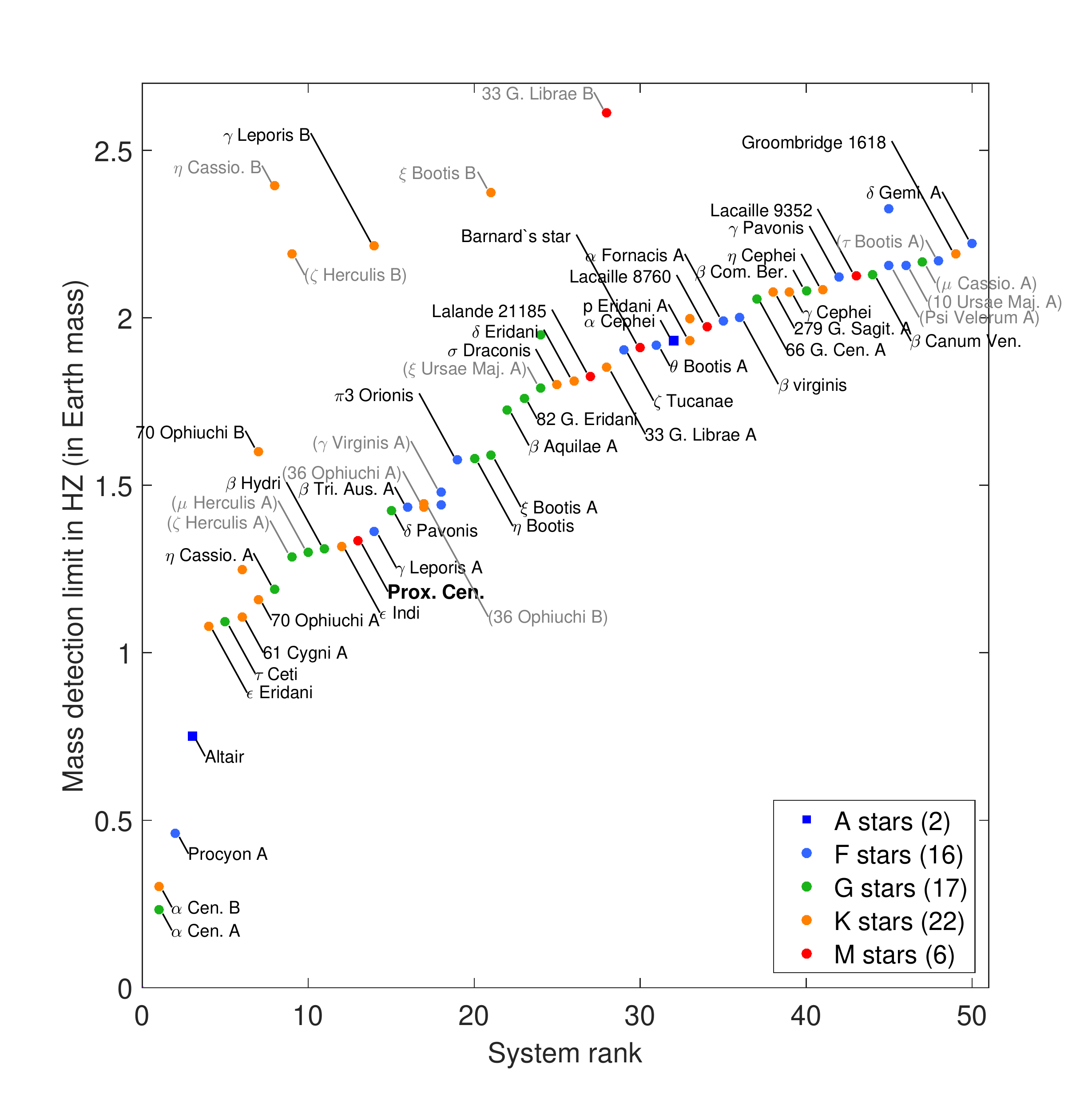}
\caption{\em Minimum masses of planets that can be detected at the center of the HZ of their star for the 63 best nearby A, F, G, K, M target systems. The target systems (either single or binary stars), are ranked from left to right with increasing minimum detectable mass in HZ around the primary system component, assuming equal observing time per system. Thus for binary stars, A and B components are aligned vertically, as they belong to the same system they share the same rank. When the A and B mass thresholds are close the name is usually not explicitly written down to avoid overcrowding. B components that have mass thresholds above 2.2 $M_\oplus$ are named in gray and binaries that are estimated too close for follow-up spectroscopy are named in gray and in parenthesis. These binaries are expected to be only part of the secondary science program (planet formation around binaries). The star sample that is best for astrometry is similar to that of the best stars for spectroscopy in the visible, or in thermal IR (see text for explanations). Earths and super-Earths with $M_{\rm p} \geq 1.5$ $M_\oplus$ can be detected and characterized (actual mass and full orbit) around 22 stars. All Super-Earths with $M_{\rm p} < 2.2$ $M_\oplus$ can be detected and characterized around 59 stars.
}
\label{fig:exoplan1}
\end{figure*}

\begin{figure*}[tb]
\centering
\includegraphics[width=0.95\textwidth]{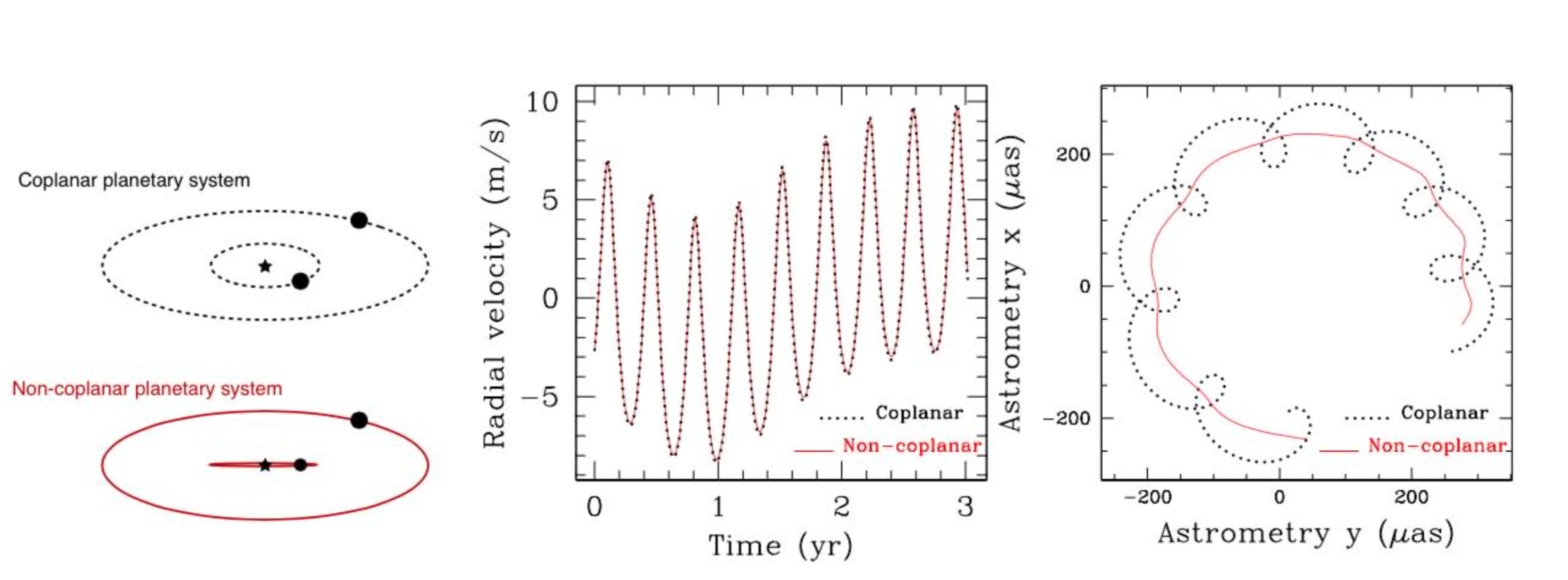}
\caption{\em An example where astrometry breaks the degeneracy. Two simulated planetary systems are around a solar-type star at 10 pc, with two Jupiter-like planets at 0.5 and 2.5 AU (\textit{left}). One is co-planar (\textit{dotted black line}), the other has a mutual inclination of 30$^\circ$ (\textit{full red line}). The two corresponding RV curves are shown (\textit{middle}), as well as the two astrometric ones (\textit{right}). Curves are identical in the former case, but clearly separated in the latter revealing the inclined orbits.}
\label{fig:exoplan2}
\end{figure*}

Surgical single-point positional precision measurements in pointed, differential astrometric mode ($<1\mu$as), could exploit a high precision astrometric  mission's unique capability to search for the nearest Earth-like planets to the Sun. The amplitude $\alpha$ of the astrometric motion of a star due to an orbiting planet is (in micro-arcseconds):
\begin{equation}
\alpha = 3\left(\frac{M_{\rm p}}{M_\oplus}\right)\left(\frac{a_{\rm p}}{1\,\mathrm{AU}}\right)\left(\frac{M_\star}{M_\odot}\right)^{-1}\left(\frac{D}{1\,\mathrm{pc}}\right)^{-1}\,\,\mu\mathrm{as}
\end{equation}
where $M_\star$ is the stellar mass, $M_{\rm p}$ is the mass of the planet, $a_{\rm p}$ is the semi-major axis of the orbit of the planet, and $D$ is the distance to the star. For a terrestrial planet in the HZ of a nearby sun-like star, a typical value is 0.3 $\mu$as (an Earth at 1.0 AU of a Sun, at 10 pc).  This very small motion (the size of a coin thickness on the Moon as measured from the Earth) will be accessible to a high precision astrometric instrument by measuring the differential motion of the star with respect to far-away reference sources.

A core exoplanet program could be comprised of 63 of the nearest A, F, G, K, and M stars (Fig. \ref{fig:exoplan1}). Many of them are found in binary and multiple systems. Binary stars are compelling for a high precision space mission for a number of reasons.  They are easier targets than single stars. For close Sun-like binaries, the magnitude of both components is lower
 than $V=9$ mag, which is the equivalent magnitude of a typical reference star field composed of 6 $V=11$ mag stars.

 Furthermore, as the photon noise from the references is the dominant factor of the error budget, the accuracy for binaries increases faster with telescope staring time than around single stars. For binaries, the references only need to provide the plate scale and the reference direction of the local frame, the origin point coordinates are constrained by the secondary/primary component of the binary.  Finally, when observing a binary, the astrometry on both components is obtained simultaneously: the staring time is only spent once as both components are within the same FoV. These two effects combined cause the observation of stars in binary systems to be much more efficient (as measured in $\mu$as$\times h^{-1/2}$) than that of single stars.

 We further stress that the complete census of small and nearby planets around solar-type stars is unique to high-precision astrometry.  On the one hand, Sun-like stars have typical activity levels producing Doppler noise of $\sim1$ m/s (or larger), which is still 10 times the signal expected from an Earth-analog \citep{2011arXiv1107.5325L}.  High precision space astrometry will be almost insensitive to the disturbances (spots, plages) due to stellar activity, having typical activity-induced astrometric signals with amplitude below 0.1 $\mu$as \citep{2011A&A...528L...9L}.

 For the full sample of the nearest stars considered in Fig.~ \ref{fig:exoplan1} we achieve sensitivity (at the $6-\sigma$ level) to planets with $M_p \leq 3$ $M_\oplus$ (See section 3.6). If we consider $\eta_\mathrm{Earth}\sim10\%$, for the sample of 63 stars closest to our Solar System we thus expect to detect $\sim6$ HZ terrestrial planets. Of these, 5 will be amenable for further spectroscopic characterization of their atmospheres.  A high precison astrometry mission could perform the measurements
 of the relevant stars and make a  thorough census (95\% completeness) of these planets by using less than 10\% of a four years mission. 
As indicated above, this program will also be valuable for understanding planetary diversity, the architecture of planetary systems (2-d information plus Kepler's laws, 
results in 3-d knowledge) including the mutual inclination of the orbits, a piece of information that is often missing in our exploration of planetary systems. 

\subsubsection{Additioanal Exoplanet Inquiries}

A secondary  program can help elucidate other important questions in exoplanetray science.
\begin{enumerate}[label=\textbf{\alph*})]
\item {\bf Planetary systems in S-Type binary systems}. A high precision astrometry mission's performance for exoplanet detection around nearby binaries will be of crucial importance in revealing planet formation in stellar systems, the environment in which roughly half of main-sequence stars are born. The discovery of giant planets in binaries has sparked a string of theoretical studies, aimed at understanding how planets can form and evolve in highly perturbed environments \citep{2015pes..book..309T}. Giant planets around one component of a binary (S-type orbits) have often been found in orbits very close to theoretical stability limits \citep[e.g.][]{2004AIPC..713..269H, 2011CeMDA.111...29T, 2016AN....337..300S}, and as for most of the binary targets the HZ of each component is stable, finding other and smaller bodies in their HZs is a real possibility. The contribution of a high precision astrometric mission could be decisive for these ongoing studies, by allowing the exploration of a crucial range of exoplanetary architectures in binaries.
\item {\bf Follow-up of known Doppler systems}. Another unique use of a high precision astrometric mission will be the study of non-transiting, low-mass mul\-ti\-ple-planet systems that have already been detected with RVs. High precision astrometry will confirm or refute controversial detections, remove the $\sin i$ ambiguity and measure actual planetary masses. Furthermore, it will directly determine mutual inclination angles, which are critical to study (i) the habitability of exoplanets in multiple systems, since they modify the orientation of the spin axes and hence the way the climates change across time \citep[e.g.][]{1993Natur.361..608L,2013MNRAS.428.1673B,2014MNRAS.444.1873A} and (ii) the dynamical evolution history of multiple systems, as e.g. coplanar orbits are indicative of smooth evolution, while large mutual inclinations and eccentricities point toward episodes of strong interactions, such as planet-planet scattering. Fig.~\ref{fig:exoplan2} illustrates a case where degeneracy in RV can be removed by astrometry.
\item {\bf Planetary systems on and off the main sequen\-ce}. \gaia has the potential to detect thousands of giant planetary companions around stars of all ages (including pre- and post-main-sequence), spectral type, chemical abundance, and multiplicity \citep{2008A&A...482..699C,2014MNRAS.437..497S,2014ApJ...797...14P,2015MNRAS.447..287S}. 
A high precision astrometriy mission could  cherry-pick on \gaia discoveries and identify systems amenable to follow-up to search for additional low-mass components in such systems, particularly in  the regime of stellar parameters difficult for radial velocity work like early spectral types, young ages, very low metallicity, white dwarfs. Some of the systems selected 
might also contain transiting companions identified by \textit{TESS} and \textit{PLATO} (and possibly even \gaia itself), or planets directly imaged by \textit{SPHERE} or E-ELT. 
\item {\bf Terrestrial planets around Brown Dwarfs}. So far, among the few planetary mass objects that have been associated to brown dwarf (BD) hosts using direct imaging and microlensing techniques, only one is likely to be a low-mass planet \citep[][and references therein)]{2015ApJ...812...47U}.  However, there are both observational \citep{2008ApJ...681L..29S,2012ApJ...761L..20R,2014ApJ...791...20R} as well as theoretical \citep{2007MNRAS.381.1597P,2013ApJ...774L...4M} reasons to believe that such systems could also be frequent around BDs. The recent identification of a trio of short-period Earth-size planets transiting a nearby star with a mass only $\sim10\%$ more massive than the Hydrogen-burning limit \citep{2016Natur.533..221G} is a tantalizing element in this direction. In its all-sky survey, \gaia will observe thousands of ultra-cool dwarfs in the backyard of the Sun with sufficient astrometric precision to reveal any orbiting companions with masses as low as that of Jupiter \citep{2014MmSAI..85..643S}. A high precision astrometry mission could push detection limits of companions down to terrestrial mass. If the occurrence rate of $P\leq1.3$ d, Earth-sized planets around BDs is $\eta=27\%$ as suggested by \citet{2017MNRAS.464.2687H} based on extrapolations from transit detections around late M dwarfs, the high precision measurements, probing for the first time a much larger range of separations with respect to transit surveys with sensitivity to low-mass planets, will unveil a potentially large number of such companions, and place the very first upper limits on their occurrence rates in case of null detection.
\end{enumerate}
\subsection{Compact objects}
\label{sec:compact-objects}

%

\subsubsection{X-ray Binaries}
The brightest Galactic X-ray sources are accreting compact objects in binary systems.  Precise optical astrometry of these X-ray binaries provides a unique opportunity to obtain quantities which are very difficult to obtain otherwise.  In particular, it is possible to determine the distances to the systems via parallax measurements and the masses of the compact objects by detecting orbital motion to measure the binary inclination and the mass function.  With 
a high precision astrometric mission, distance measurements are feasible for $>$50 X-ray binaries (in 2000h), and orbital measurements will be obtained for dozens of systems.  This will revolutionize the studies of X-ray binaries in several ways, and here, we discuss goals for neutron stars (NSs), including constraining their equation of state (EoS), and for black holes (BHs).

\begin{figure}[t]
\centering
\includegraphics[width=0.8\textwidth]{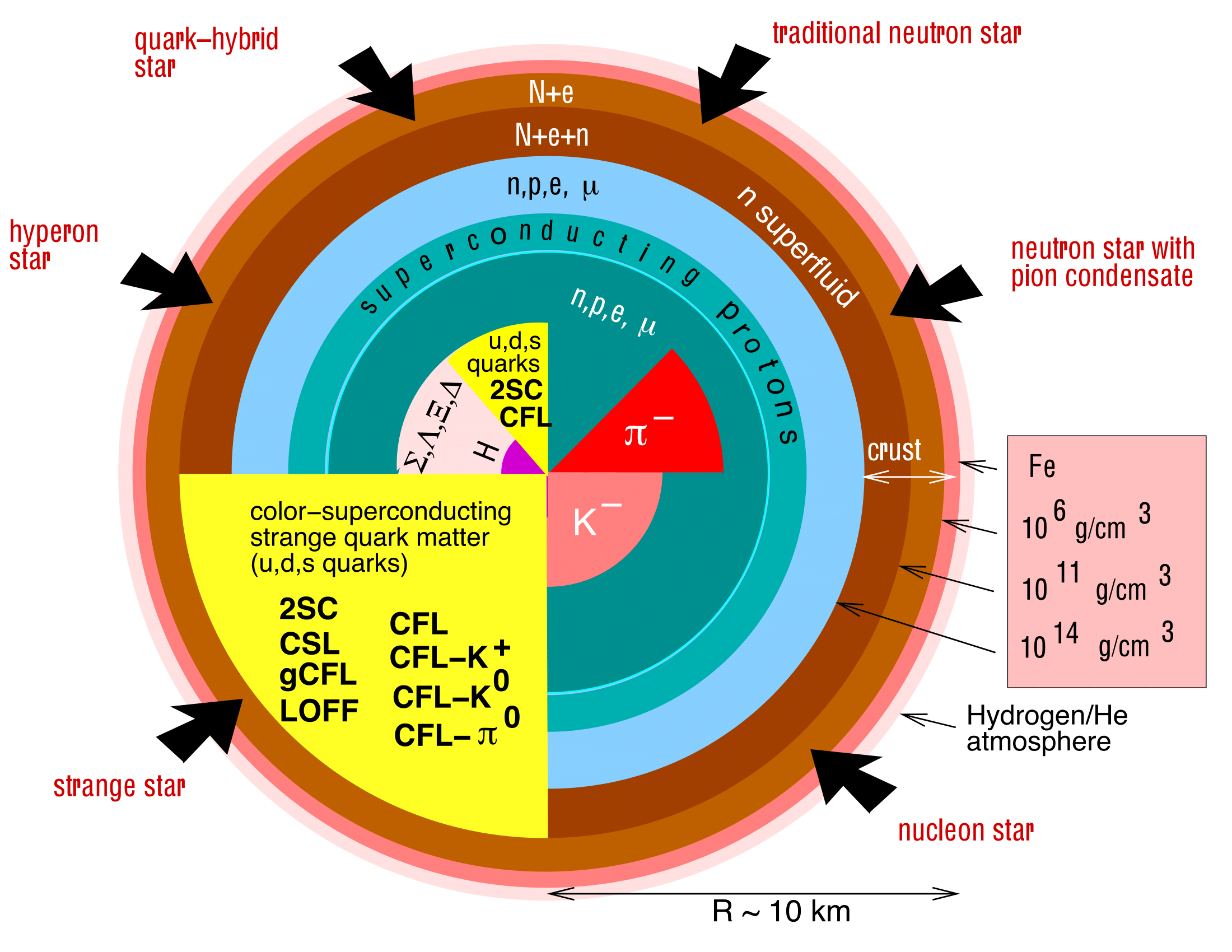}
\caption{\em Sketch of the different existing possibilities for the internal structure of a neutron star.  Figure courtesy of Fridolin Weber.}
\label{fig:nsstructure}
\end{figure}

Matter in the {\bf NS} interior is compressed to densities exceeding those in the center of atomic nuclei, opening the possibility to probe the nature of the strong interaction under conditions dramatically different from those in terrestrial experiments and to determine the NS composition.  NSs might be composed of nucleons only, of strange baryons (hyperons) or mesons in the core with nucleons outside (a hybrid star), or of pure strange quark matter (a quark star).  A sketch of the different possibilities is given in Fig.~\ref{fig:nsstructure}. Via the equation of state (EoS), matter properties determine the star's radius for a given mass. In particular, since general relativity limits the mass for a given EoS, the observation of a massive NS can exclude EoS models. Presently, the main constraint stems from the measurements of two very massive NSs in radio pulsar/white dwarf systems which have been reported with high precision \citep{Demorest2010, Antoniadis2013, Fonseca2016}.

\begin{figure*}[htb]
\includegraphics[width=0.35\textwidth]{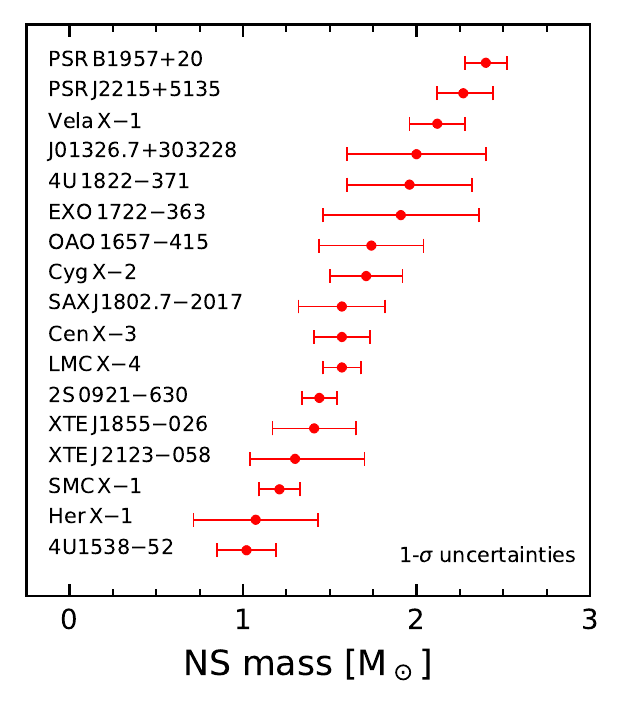}
\includegraphics[width=0.45\textwidth]{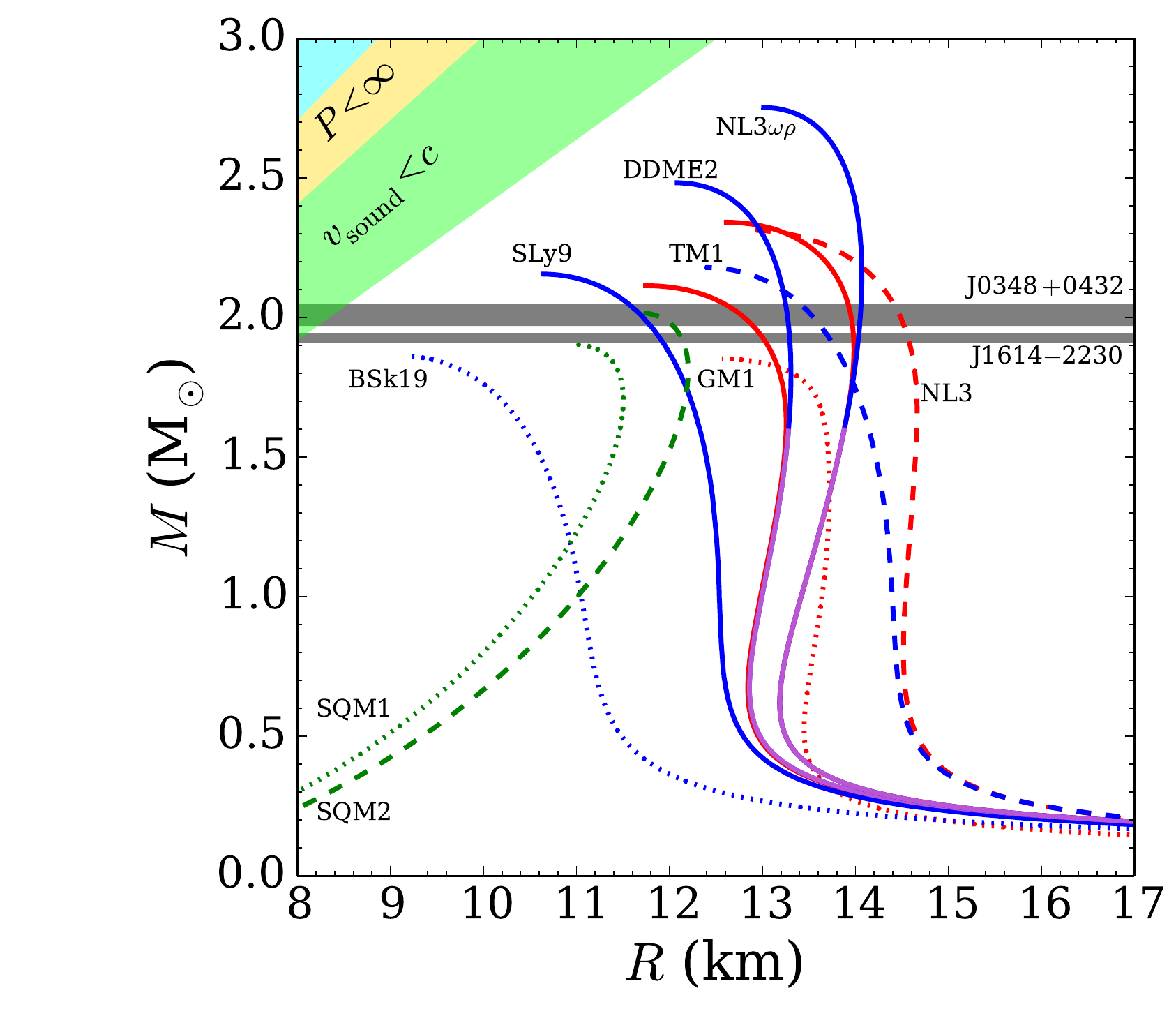}
\caption{\em Left: Neutron star mass measurements in X-ray binaries, update from \citet{Lattimer2005}, \url{http://stellarcollapse.org}.  Right: $M$-$R$ relation for different EoS models (adapted from \citet{Fortin2016}): NS with a purely nucleonic core (in blue), with a core containing hyperons at high density (in red), and pure strange quark stars (in green). The horizontal grey bars indicate the masses of PSR J1614$-$2230 and PSR J0348+0432. The models indicated by dotted or dashed lines are either not compatible with NS masses or nuclear physics constraints. Note that a transition to matter containing hyperons is not excluded by present constraints.}
\label{fig:nsmasses}
\end{figure*}

The key to constraining the NS EoS is to measure the masses and radii of NSs.  While masses have been measured for a number of X-ray binary and radio pulsar binary systems (e.g., \citet{Lattimer2016, Ozel2016}) , the errors on the mass measurements for most X-ray binaries are large (see Fig.~\ref{fig:nsmasses}, left).  The ultimate constraint on the EoS will be a determination of radius and mass of the same object, and a small number of such objects might be sufficient to pin down the entire EoS (e.g. \citet{Ozel2009}), see Fig.~\ref{fig:nsmasses} (right), where several $M$-$R$ relations for different EoSs are shown.  Current techniques to determine radii rely on spectroscopic measurements of accreting neutron stars, either in quiescence \citep{Heinke2014} or during thermonuclear (type I) X-ray bursts \citep{Ozel2016}, and also timing observations of surface inhomogeneities of rotating NSs \citep{Miller2016, Haensel2016}.

A high precision astrometric mission will contribute by obtaining precise mass constraints with orbital measurements \citep{Tomsick2010} and by improving distance measurements.  Distances must be known accurately to determine the NS radii.  For that purpose, new high precision data can be combined with existing and future X-ray data, e.g., from \textit{Athena}, which is scheduled as an ESA L2 mission.  The \textit{Athena} Science Working Group on the endpoints of stellar evolution has observations of quiescent neutron star X-ray binaries to determine the NS EoS as its first science goal; however, their target list is restricted to systems that are in globular clusters.  A high precision astrometric mission will enable distance measurements for many more NS X-ray binaries, allowing Athena to expand their target list.

Other techniques for constraining the NS EoS might also be possible in the future: detecting redshifted absorption lines; determining the moment of inertia of the double pulsar J0737$-$3039; and the detection of gravitational wave emission from the inspiral of a NS-NS merger \citep{Abbott2017}.  However, the mass and distance measurements that a high precision astrometric mission will obtain use techniques that are already well-established, providing the most certain opportunity for greatly increasing the numbers of NSs with mass or radius determinations.

In addition to the goal of constraining the NS EoS, NS masses are also relevant to NS formation and binary evolution.  Current evolutionary scenarios predict that the amount of matter accreted, even during long-lived X-ray binary phases, is small compared to the NS mass.  This means that the NS mass distribution is mainly determined by birth masses.  Determining the masses of NSs in X-ray binaries, therefore, also provides a test of current accretion models and evolutionary scenarios, including the creation of the NSs in supernovae.

BHs are, according to the theory of general relativity, remarkably simple objects.  They are fully described by just two parameters, their mass and their spin. Precise masses are available for very few BHs in X-ray binaries and the recent detection of gravitational waves \citep{Abbott2016c} found in the binary BH mergers that have been detected in gravitational waves \citep{Abbott2016b,LIGO18} have, on average, higher masses and probably lower spins than the BHs in X-ray binaries.  These measurements are difficult to explain based on our understanding of stellar evolution and the fate of massive stars. Although BHs leave few clues about their origin, one more parameter that can be determined is the proper motion of BHs in X-ray binaries. Measurements of proper motions provides information about their birthplaces and formation. It includes whether they were produced in a supernova (or hypernova) or whether it is possible for massive stars to collapse directly to BHs.  A few BH X-ray binaries have proper motion measurements (e.g., \citet{Mirabel2001}), but this number will rise dramatically with the astrometry measurements that a high precision astrometry mission will provide.

Currently, the cutting edge of research in BH X-ray binaries involves constraining BH spins, including the rate of spin and the orientation of the spin axis.  Techniques for determining the rate of spin include measuring of the relativistic broadening of the fluorescent iron $K_\alpha$ line in the X-ray emission and the study of the thermal continuum X-ray spectra |\citep{Remillard2006, Miller2007}. Concerning the direction of their spin axes, there is evidence that the standard assumption of alignment between the BH spin and orbital angular momentum axes is incorrect in some, if not many, cases \citep{Maccarone2002, Tomsick2014, Walton2016}, likely requiring a warped accretion disc. Theoretical studies show that such misalignments should be common \citep{King2016}. However, binary inclination measurements rely on modeling the ellipsoidal modulations seen in the optical light curves \citep{Orosz2011}, which is subject to systematic uncertainties, and a high precision astrometry mission will be able to provide direct measurements of orbital inclination for many of the BH X-ray binaries that show evidence for misalignments and warped discs.

\subsubsection{Astrometric microlensing}
\label{sec:Compact objects in the GC}

In 1986  Bohdan Paczy{\'n}ski \citep{Paczynski1986} proposed a new method for finding compact dark objects, via photometric gravitational microlensing. This technique relies on continuous monitoring of millions of stars in order to spot its temporal brightening due to space-time curvature caused by a presence and motion of a dark massive object. Microlensing reveals itself also in astrometry, since the centre of light of both unresolved images (separated by $\sim$1 mas) changes its position while the relative brightness of the images changes in the course of the event. Astrometric time-series at sub-mas precision over the course of a couple of years will provide measurement of the size of the Einstein Ring, which combined with photometric light curve, will directly yield the lens distance and mass. Most microlensing events are detected by large-scale surveys, e.g., OGLE and, in future possibly also the LSST.  At typical brightness of V=19-20mag only
a high-precision astrometry mission will be capable at providing good-enough astrometric follow-up of photometrically detected microlensing events. Among 2000 events found every year, at least a couple should have a black hole as the lens, for which the mass measurement via astrometric microlensing will be possible.
\begin{figure*}[htb]
  \centering
  \includegraphics[width=.98 \textwidth]{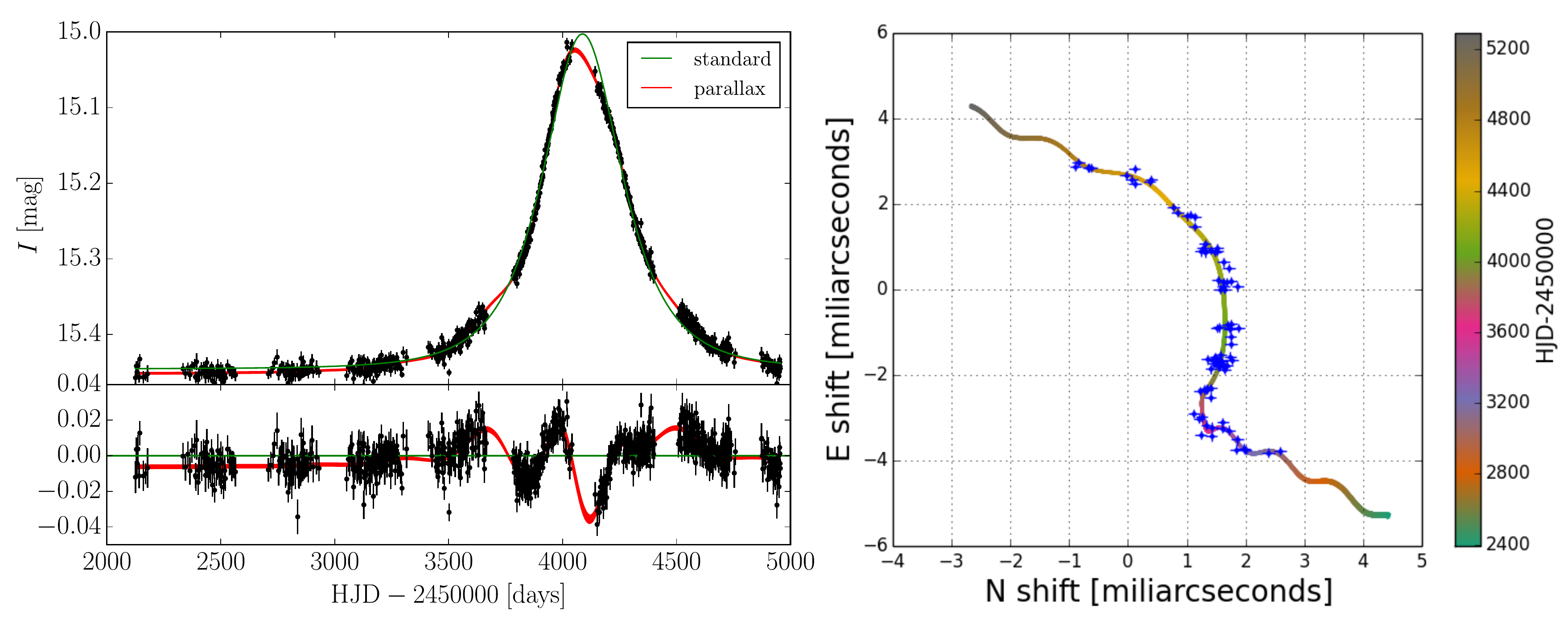}
  \caption{\em Microlensing event, OGLE3-ULENS-PAR-02, the best candidate for a $\sim$10M$_\odot$ single black hole. Left: photometric data from OGLE-III survey from 2001-2008. Parallax model alone can only provide mass measurement accuracy of 50-100$\%$. Right: simulated astrometric microlensing path for a similar event if observed with \theia. Combining superb a high precision astrometry mission's astrometric accuracy with long-term photometric data will yield mass measurements of black holes and other dark compact object to 1$\%$ even at faint magnitudes.}
\end{figure*}

\begin{figure*}[t] 
\centering
\includegraphics[width=0.75\textwidth]{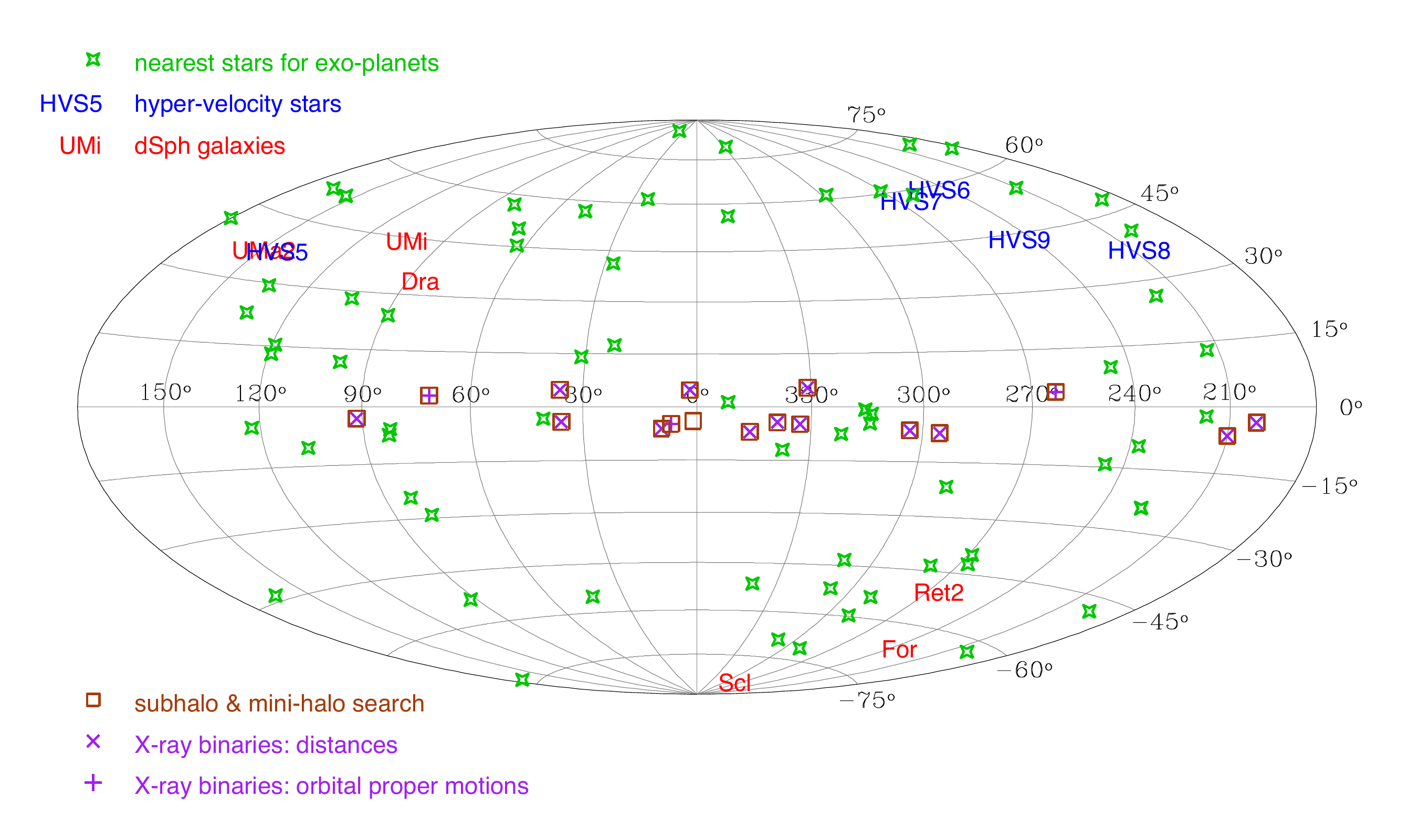}
\caption{\em Sky map of the  targets considered for observations with a high precision astrometric mission.} 
\label{fig:skyplot}
\end{figure*}

Detection of isolated black holes and a complete census of masses of stellar remnants will for the first time allow for a robust verification of theoretical predictions of stellar evolution. Additionally, it will yield a mass distribution of lensing stars as well as hosts of planets detected via microlensing. 

\subsection{Cosmic distance ladder}
\label{sec:cosm-dist-ladd}

The measure of cosmological distances has revolutionized modern cosmology and will continue to be a major pathway to explore the physics of the early Universe. The age of the Universe ($H_0^{-1}$) is a key measurement in non-standard DM scenarios.  Its exact value is currently strong\-ly debated, with a number of scientific papers pointing at discrepancies in between measurements methods at the 2-3$\sigma$ level. But the most serious tension appears between CMB estimates ($H_0 = 67.8 \pm 0.9$\,km/s/Mpc) or for that matter BAO results from the SDSS-III DR12 data, combined with SNIa which indicate $H_0 = 67.3 \pm 1.0$\,km/s/ Mpc \citep[see][]{Alam2016b} and measurements based on Cephe\-ids and SNIa ($H_0 = 73.24 \pm 1.74$\,km/s/Mpc) with a discrepancy at the 3-4 $\sigma$ level.

The tension between the methods can be due to unknown sources of systematics, to degeneracies between cosmological parameters, or to new physics \citep[e.g.][]{Karwal&Kamionkowski16}. It is therefore of crucial importance to consider methods capable of measuring $H_0$ with no or little sensitivity to other cosmological parameters. Uncertainties can be drastically reduced by measuring time delays (TD) in gravitationally lensed quasars \citep{Refsdal64}, as this technique only relies on well-known physics (GR). With enough statistics, and a good modeling the mass distribution in the lensing galaxy, TD measurements can lead to percent-level accuracy on $H_0$, independently of any other cosmological probe (e.g. \citet{Bonvin2016a, Suyu2013, Suyu2014}.  In practice, TDs can be measured by following the photometric variations in the images of lensed quasars. As the optical paths to the quasar images have different lengths and they intersect the lens plane at different impact parameters, the wavefronts along each of these paths reach the observer at different times. Hence the notion of TD.

Significant improvements in lens modeling, combined with long-term lens monitoring, should allow measuring $H_0$ at the percent level. The H0LiCOW program ($H_0$ Lenses in COSMOGRAIL's Wellspring), which focuses on improving the detailed modeling of the lens galaxy and of the mass along the line of the sight to the background quasar, led to $H_0 = 71.9 \pm 2.7$\,km/s/Mpc (that is 3.8\% precision) in a flat LCMD Universe by using deep HST imaging, Keck spectroscopy and AO imaging and wide field Subaru imaging \citep{Suyu2016, Rusu2016, Sluse2016, Wong2016, Bonvin2016a}. This value is in excellent agreement with the most recent measurements using the distance ladder (though in tension with the CMB measurements from Planck) but still lacks of precision.

By performing photometric measurements with the required sensitivity and no interruption, the combination of a high precision astrometric mission and excellent modeling of the lens galaxy, will enable to measure $H_0$ at the percent level and remove any possible degeneracies between $H_0$ and other cosmological parameters. This will open up new avenues to test the DM nature. An alternative technique consists in using  trigonometric parallaxes. This is  the only  (non-statistical and model-independent) direct measurement method and the foundation of the distance scale. 
A high precision astrometric mission has the potential to extend the  "standard candles" - the more distant pulsating variables: Cepheids, RR Lyrae, Miras  and  also Stellar Twin stars -   well beyond the reach of \gaia. 

These distance measurements can be transferred to nearby galaxies allowing us to convert observable quantities, such as angular size and flux,  into physical qualities  such as energy and luminosity. Importantly, these  distances scale linearly with $\hn$, which gives the temporal and spatial scale of the universe. With this improved knowledge, we will then be able to to better understand the structure and evolution  of both our own and more distant galaxies, and the {\it age} of our universe.

\subsection{Position of the science targets in the sky}
\label{sec:posit-science-tar}

The different targets considered for observations with a high precision astrometry mission have been located in Fig.~\ref{fig:skyplot} on a sky map.

\section{Possible space mission}
\label{sec:poss-space-miss}

Several mission profiles have been considered in the last few years focused in differential astrometry, for instance NEAT, micro-NEAT and \theia. Additional new differential astrometry mission configurations adapted with technological innovations will certainly be envisioned to pursue accurate measurements of the extremely small motions required by the science cases in this white paper.  

\subsection{Scientific requirements}
\label{sec:scient-requ}

To address the science described in this white paper, a high precision astrometry mission should stare towards :
\begin{itemize}
\setlength\itemsep{0em} 
\item dwarf galaxies (Sphs), to probe their DM inner structure;  
\item hyper-Velocity stars (HVSs), to probe the triaxiality of the halo, the existence of mini compact halo objects and the time delay of quasars; 
\item the Galactic disc, to probe DM subhalos and mini compact halo objects; 
\item star systems in the vicinity of the Sun, to find the nearest potentially habitable terrestrial planets;
\item known X-ray binaries hosting neutron stars or Black Holes.
\end{itemize}

\begin{figure}[t]
\centering
\includegraphics[width=0.8\textwidth,trim = 0.8cm 0cm 0cm 0cm, clip]{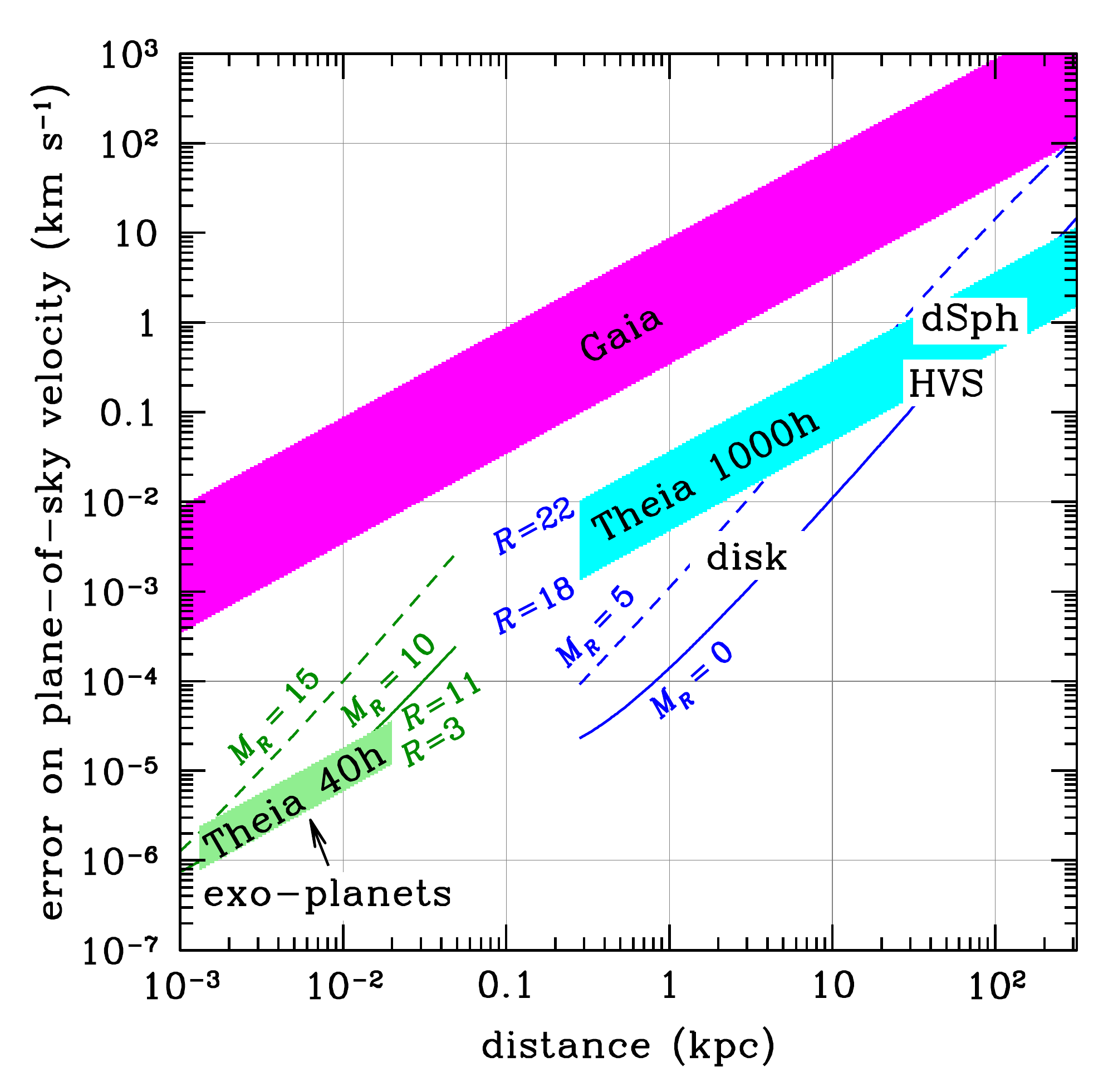}
\caption{\em Expected plane-of-sky velocity errors from a high precision astrometry mission's proper motions as a function of distance from Earth. These errors respectively correspond to 40 and 1000 cumulative hours of exposures for exo-planets (\textit{green}) and more distant objects (\textit{cyan} and \textit{blue}),  during a 4 year interval for observations, including the systematic limit from calibration on \gaia reference stars. The expected precision for specific objects are highlighted. The accuracy  for the 5-year \gaia mission is shown in \textit{magenta}.}
\label{fig:vposerrvsDist}
\end{figure}

\begin{figure*}[t] 
\centering
\includegraphics[width=0.42\hsize,clip]{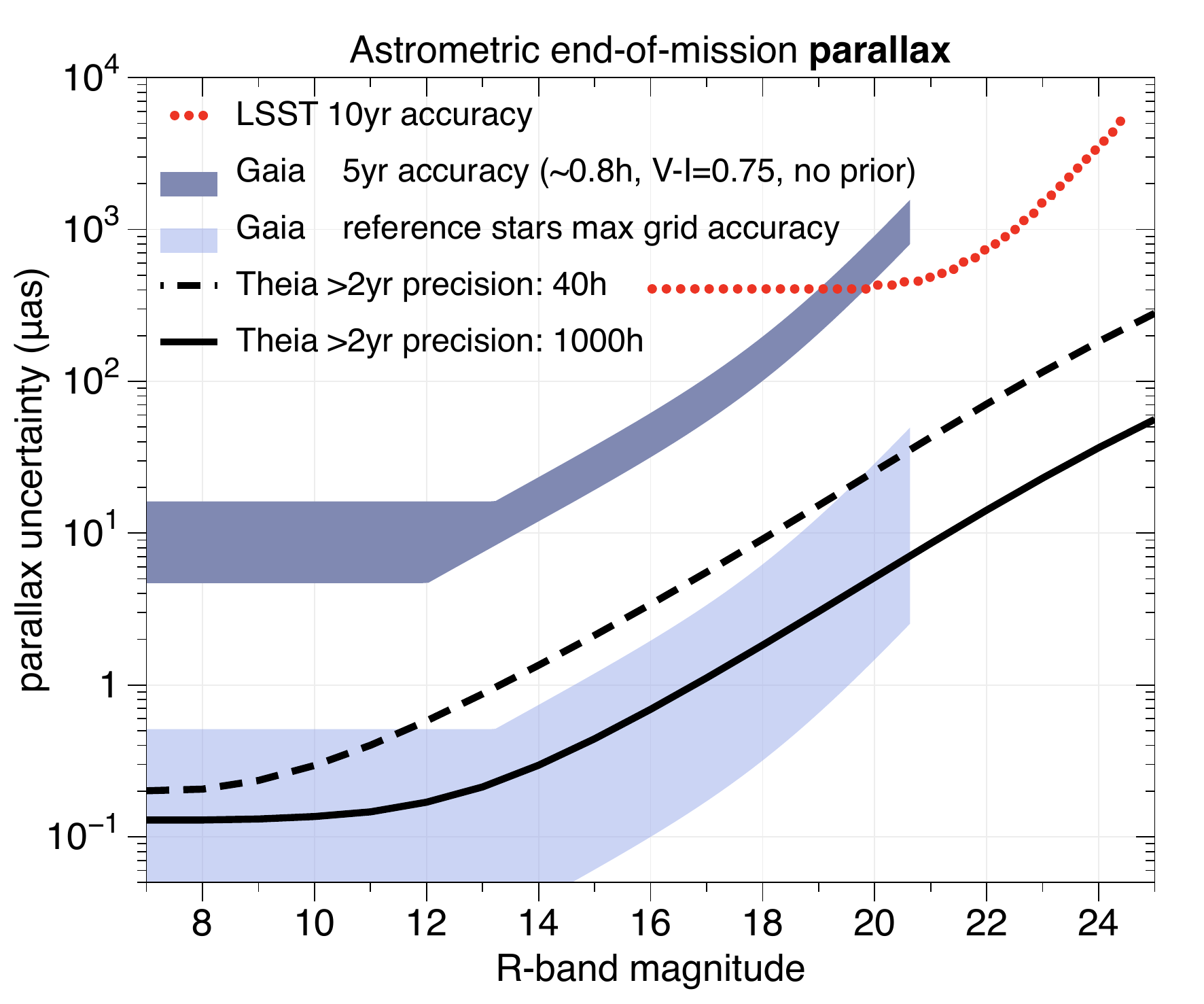}
\includegraphics[width=0.42\hsize,clip]{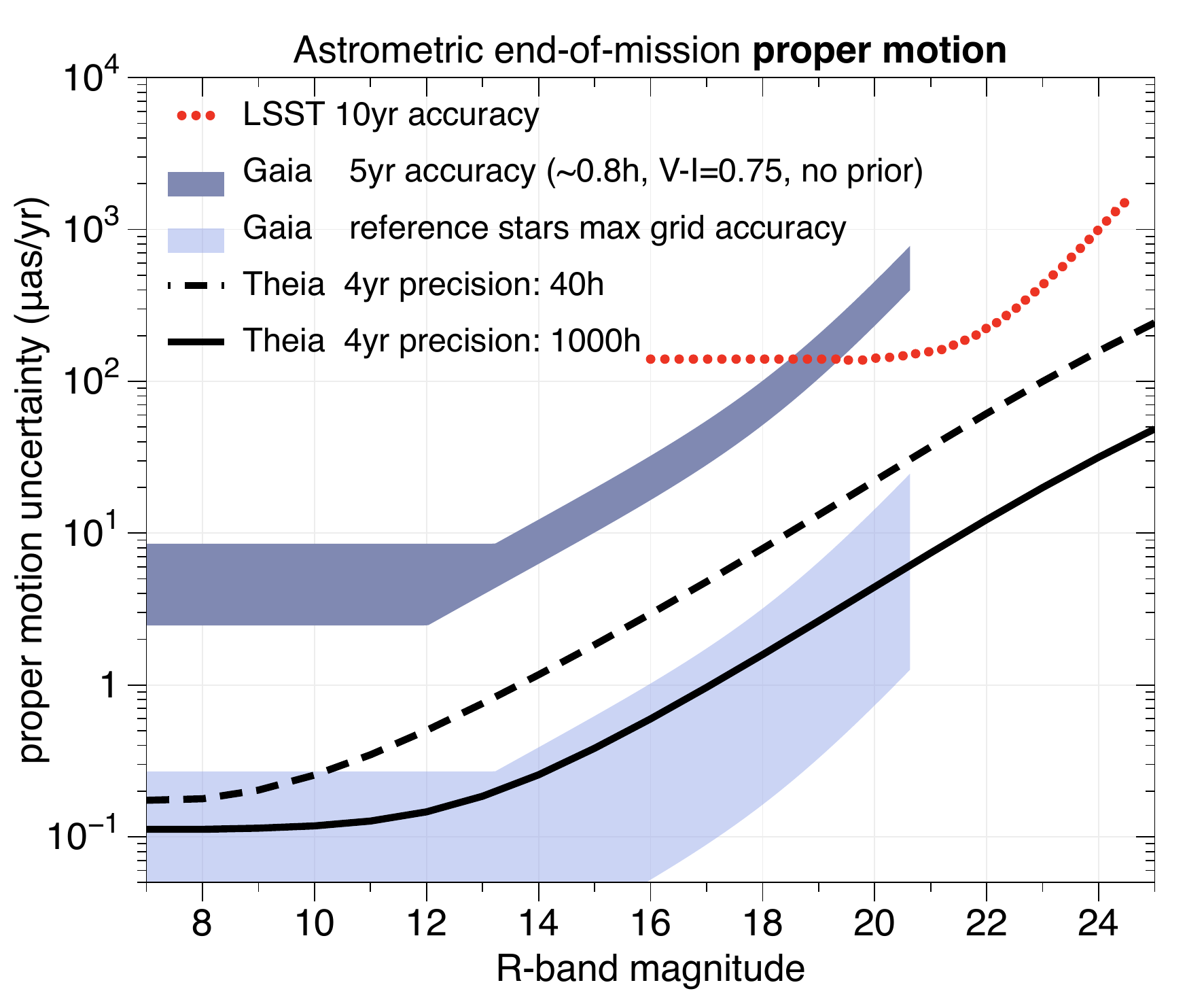}
\caption{Estimated RMS precision on  a high precision astrometry missionrelative parallax (\textit{left}, for ecliptic latitude $0^\circ$) and proper motion (\textit{right}) in the $R$-band. Also shown for comparison are the estimated accuracies for 10~years LSST \citep{2009arXiv0912.0201L} as well as the 5-year nominal \gaia mission \citep{deBruijne+15} (vertical spread caused by position on the sky, star colour, and bright-star observing conditions). Small-scale spatial correlations ($<$1$^\circ$) between \gaia reference sources will limit the maximum reachable absolute parallax and proper motion calibration for a high precision astrometry mission, indicated by the light blue band for a range of assumed spatial correlations as function of reference star magnitude.}
\label{fig:astroComparisons}
\end{figure*}

\begin{table*}[t]
  \small
  \caption{\em Summary of science cases with most stringent performance requirements set in each case. Figures are based on a 4 year mission, thermal stabilisation ($+$slew time) is assumed to take 30$\%$ of the mission time.
  \label{tab:tech.summary.science}} 
  \centering
  \begin{tabular}{lrrrrr} 
    \hline 
Program 
& Used
& Mission 
& Nb of objects
& Benchmark target 
& EoM precision \\
& time (h)
& fraction 
& per field 
& $R$ mag (and range) 
& (at ref.\ mag.) \\
    \hline 
    \hline 
    Dark Matter & 
    17\,000 &
    0.69 &
    10$^2$--10$^5$ &
    20 (14--22) &
    10 $\mu$as  \\ 
    $\&$ compact objects  
    \\
    \hline
    Nearby Earths           
    & 3\,500 &
    0.14 &
    $<$20 &
    5 (1--18)  &
    0.15 $\mu$as  \\ 
    $\&$ follow-up \\
    \hline
    Open observatory & 
    4\,000 &
    0.17 &
    10-10$^5$ &

    6 (1-22) &
    1.0 $\mu$as  \\
    \hline 

    Overall requirements& 
    24 \,500 &
    1.00 &
    10$^1$-10$^5$ &
    6 (1-22) &
    0.15-10 $\mu$as \\  
    \hline 
  \end{tabular}
\end{table*}
For a targeted mission, the objects of interest must be sampled throughout the lifetime of the mission. After re-pointing the telescope and while waiting for stabilization, photometric surveys, e.g. for measurements of $H_0$ using lensed quasar time delays could be performed, thus optimizing the mission scientific throughput. Fig.~\ref{fig:skyplot} shows a sample sky map with potential targets. 

As illustrated in Fig.~\ref{fig:vposerrvsDist}, high precision astrometric missions could measure the plane-of-sky velocities of the faintest objects in the local Universe, with errors as small as a few mm/s in the case of the hosts of Earth-mass exo-planets in the habitable zone of nearby stars, a few m/s for stars in the Milky Way disc, i.e. for kinematical searches for DM sub-halos, micro-lensing searches for ultra-compact mini-haloes, and for the companions of neutron stars and black holes in X-ray binaries, 200m/s for hyper-velocity stars whose line of sight velocities are typically $>500$km/s, and finally 1km/s for $R=20$ stars for dwarf spheroidal galaxies.

A mission concept with an expected \theia-like astrometric precision, as shown in Fig.~\ref{fig:astroComparisons}, surpasses what will be achieved by other approved space astrometric surveys and ground surveys, thus unlocking science cases that are still unreachable.

Table \ref{tab:tech.summary.science} summarizes the science cases with most stringent performance requirements. To cover the science questions from this white paper, any mission concept must be flexible, allowing for observing modes covering a wide flux dynamical range. This requires the concepts to cope with \emph{Deep Field Modes}, aimed towards objects as dwarf galaxies, and \emph{Bright Star Modes}, focused in the study of planetary systems around nearby stars.

\subsection{Example of a M-size mission}
\label{sec:prop-scient-instr}

\begin{figure*}[t]
  \centering
  \includegraphics[width=0.9\hsize]{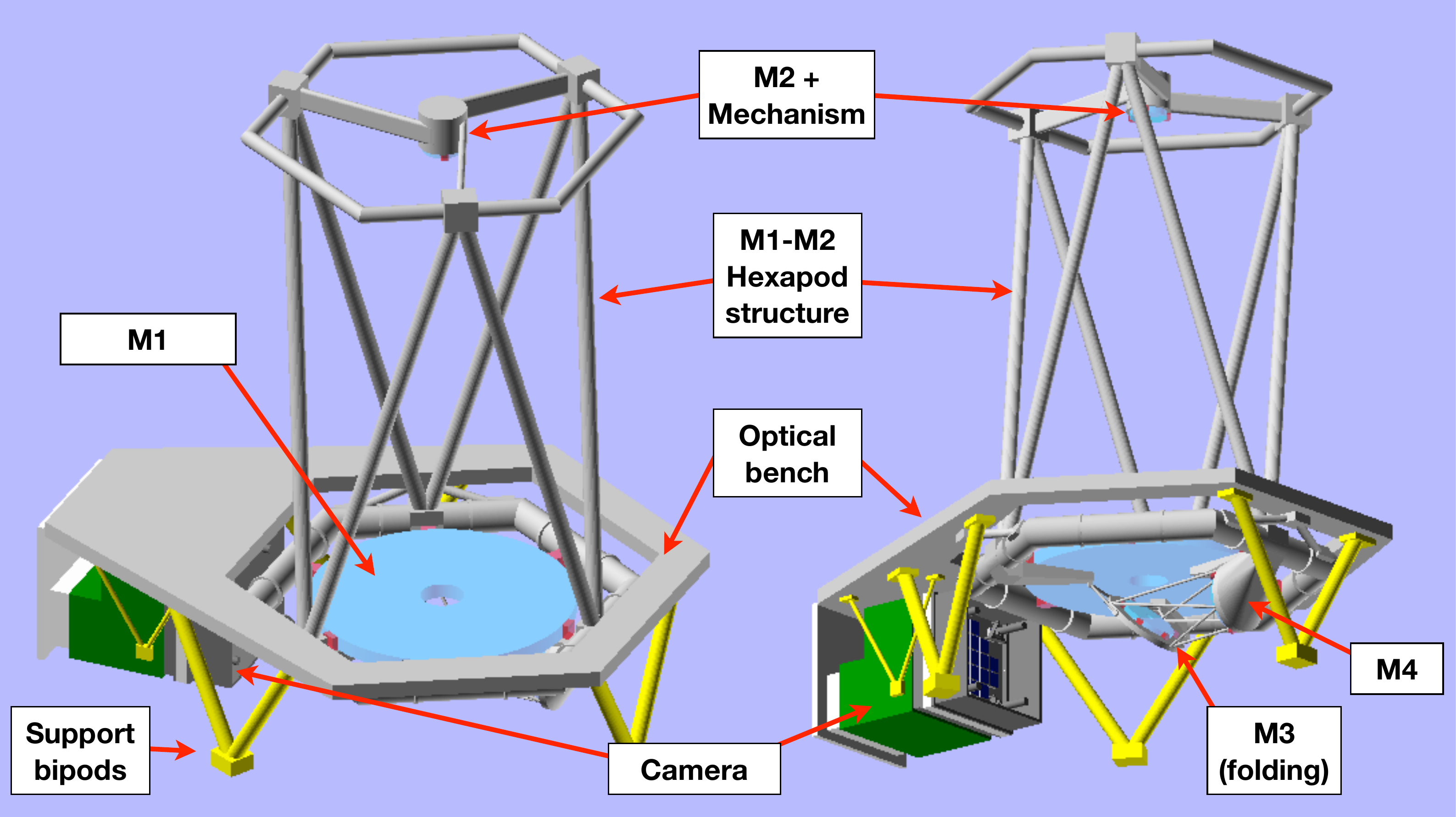}  
  \caption{\em Overall layout of the \theia Payload Module concept. Volume is estimated in $1.6\times1.9\times2.2$m$^3$.}
  \label{fig:theiam5-plmconcept}
\end{figure*}

The Payload Module (PLM) of a high precision astrometric mission must be simple. It is essentially composed from four subsystems: telescope, camera, focal plane array metrology and telescope metrology. In the case of the \theiax/M5 concept, they were designed applying heritage from space missions and concepts like \gaia, \hst/FGS, \textit{SIM}, \NEATx/M3, \theiax/M4 and \euclidx. 

However achieving micro-arcsecond differential astrometric precision requires the control of all effects that can impact the determination of the relative positions of the point spread function. The typical apparent size of an unresolved star corresponds to 0.2~arcseconds for a 0.8~m telescope operating in visible wavelengths. The challenge is therefore to control systematics effects to the level of 1 part per 200\,000. The precision of relative position determination in the Focal Plane Array (FPA) depends on i) the photon noise, which can be either dominated by the target or by the reference stars; ii) the geometrical stability of the instrument, iii) the stability of the optical aberrations, iv) the variation of the detector quantum efficiency between pixels. The control of these effects impairs other missions that otherwise could perform micro-arcsecond differential astrometry measurements, like \hst, {\textit Kepler}, \wfirst or \euclidx, posing fundamental limits to their astrometric accuracy. All these effects must be taken into account in any high precision differential astrometry mission concept.

To address the challenges and fulfill the requirements from section \ref{sec:scient-requ}, two different possible concepts can be investigated. A \textit{NEAT}-like mission consisting on a formation flight configuration \citep{2012ExA....34..385M} or an \euclidx-like mission,\footnote{\euclid red book: \url{http://sci.esa.int/euclid/48983-euclid} \url{-definition-study-report-esa-sre-2011-12}.} but with a single focal plane and additional metrology subsystems. Both concepts consist in adopting a long focal length, diffraction limited, telescope and additional metrological control of the focal plane array. The proposed \theia/M5 mission concept was the result of a trade-off analysis between both concepts. 

\subsubsection{Telescope concept}

\begin{figure}[t]
\centering
\includegraphics[width = 0.8\hsize]{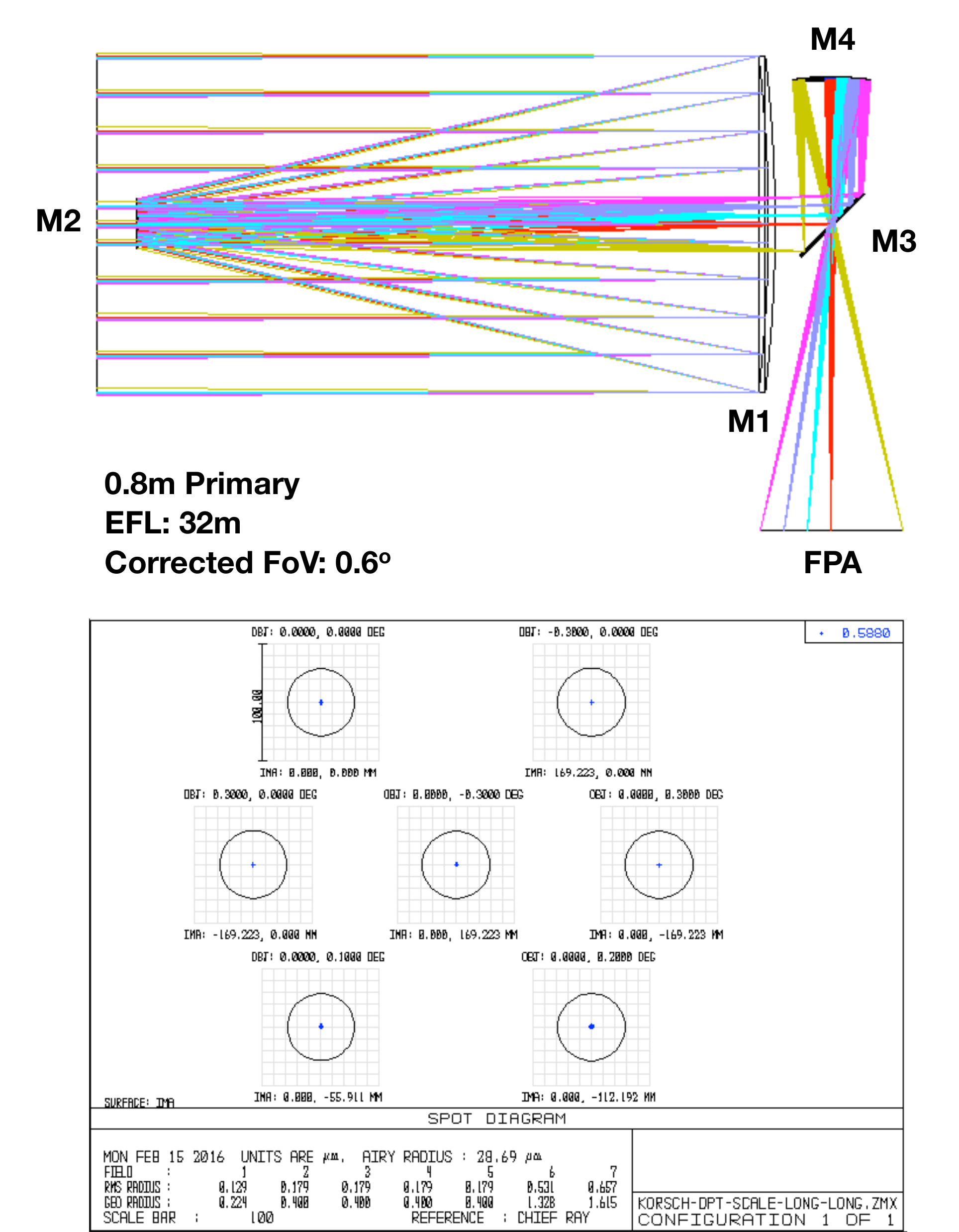}\\*[1em]
\vspace{-0.4cm}
\caption{\em On-axis Korsch TMA option. Raytracing and spot diagrams for the entire FoV. This design was adopted as the baseline for the \theia/M5 proposal.}
\label{fig:theiam5-onaxis-korsch}
\end{figure}

\begin{figure}[t]
\centering %
\includegraphics[width = \textwidth]{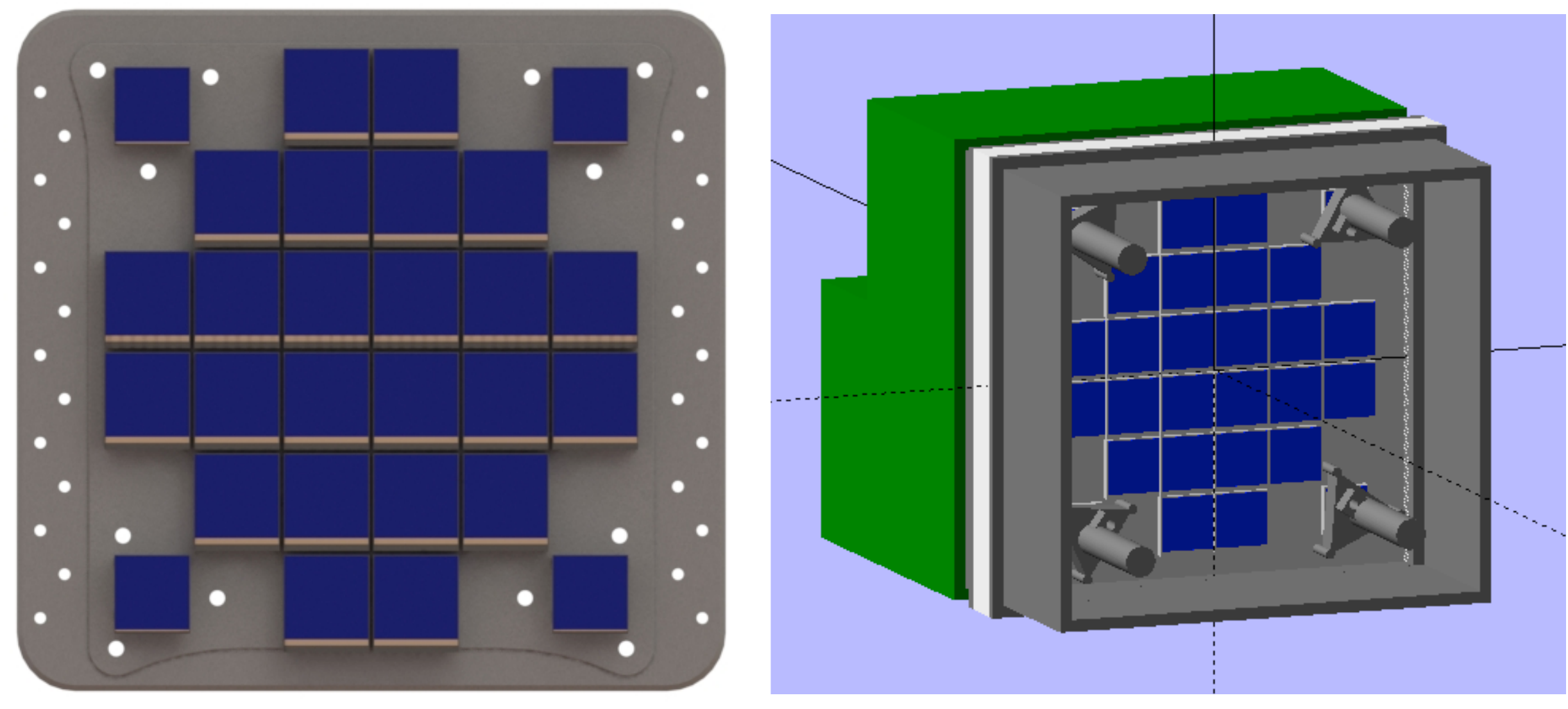}\\*[1em]
\vspace{-0.4cm}
\caption{\em Concept for the \theia/M5 Camera. Concept for the FPA detector plate at the left. Overall view of the camera concept on the right.}
\label{fig:theiam5-fpaconcept}
\end{figure}

\begin{table}[t]
  \centering
  \caption{\em \theia's mission main characteristics.}
  \smallskip
  \includegraphics[width=\hsize]{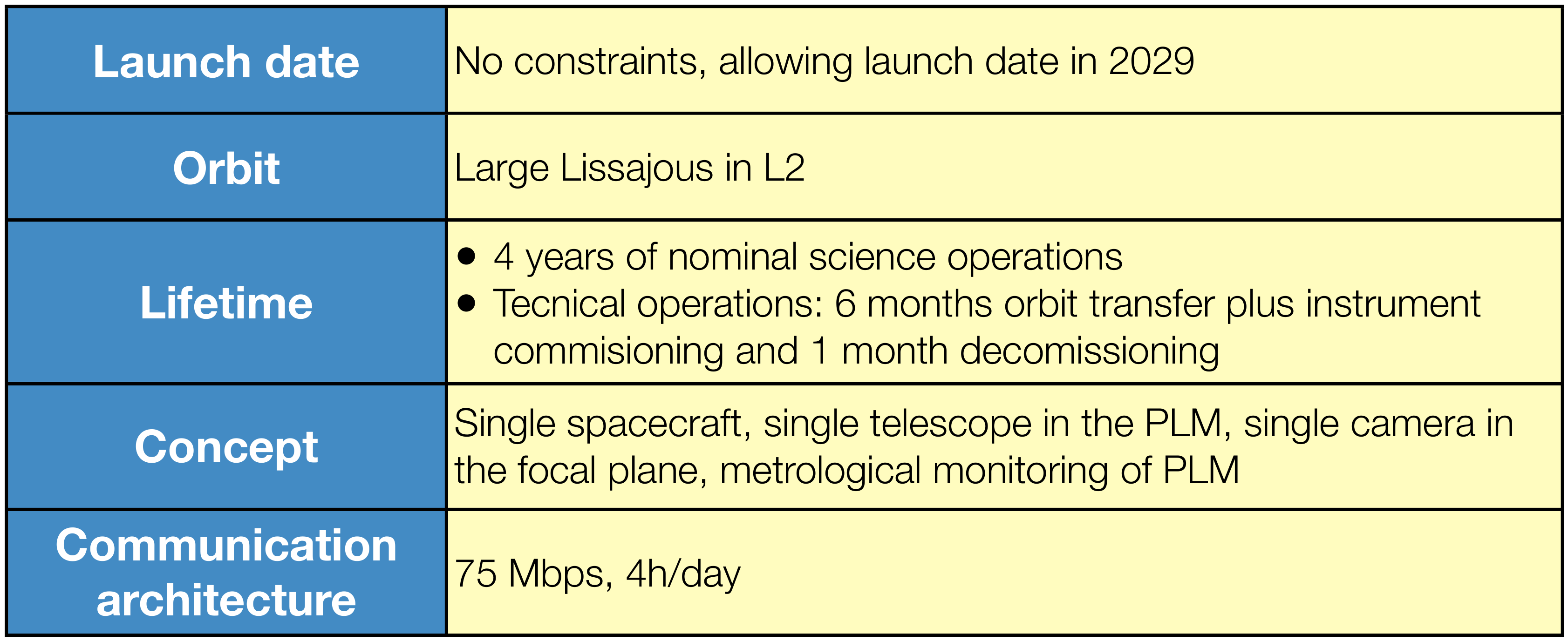}
  \bigskip
  \label{tab:mission-main-characteristics}
\end{table}

\begin{figure*}[p]
  \centering
  \includegraphics[width=0.98\hsize]{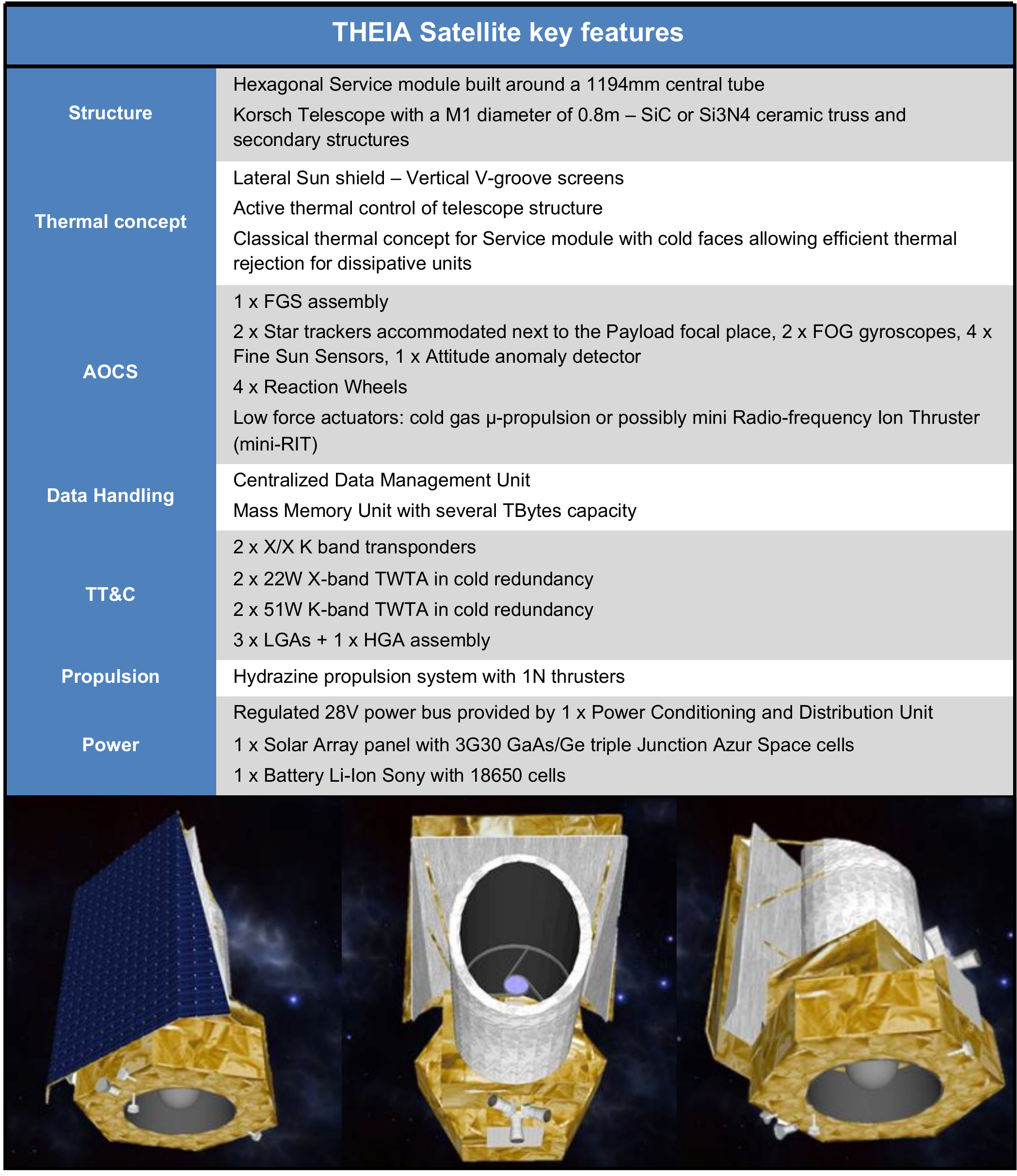}
  \caption{\em Proposed \theia satellite concept}
  \label{fig:mission.sc.concept}
\end{figure*}

The \theia PLM concept consists on a single Three Mirror Anastigmatic (TMA) telescope with a single focal plane (see Fig.~\ref{fig:theiam5-plmconcept}) covering a $0.5^\circ$ field-of-view with a mosaic of detectors. To monitor the mosaic geometry and its quantum efficiency, the PLM includes a focal plane metrology subsystem. While to monitor the telescope geometry, a dedicated telescope metrology subsystem is used.

To reach sub-microarcsecond differential astrometry a diffraction limited telescope, with all aberrations controlled, is necessary. A trade-off analysis was performed between different optical designs, which resulted in two optical concepts that could fulfill all requirements. Both are based on a Korsch Three Mirror Anastigmatic telescope; one is an on-axis solution
while the second is an off-axis telescope. In both cases only three of the mirrors are powered mirrors. While the on-axis solution adopts a single folding mirror, the off-axis solution adopts two folding mirrors. The on-axis design was the \theia/M5 baseline. More recently, however, studies from NASA/JPL show that a customized and corrected Ritchey-Chretien can reach 5 $\mu$as over a $0.5^\circ$ FoV, which even if not capable to address habitable exoplanet science cases, would provide a valuable instrument for Dark Matter studies.

To achieve the precision by centroiding as many stars as possible, a mosaic of detectors (in principle CCD or CMOS) must be assembled on the focal plane. The detectors must feature small pixels ($\sim 10 \mu$m) and well controlled systematic errors along the lifetime of the mission. Detailed in orbit calibration of the focal plane and detector geometry and response must be monitored, and in the \theia concept this is addressed via a dedicated laser metrology.

In addition to measuring the FPA, the structure of the telescope needs monitoring to control time-variable aberrations at sub $\mu$as level. Even at very stable environments such as L2 the telescope geometry varies for different reasons: structural lattice reorganization (as the micro-clanks observed in ESA/\gaia), outgassing and most importantly, thermo-elastic effects due to the necessary variation of the Solar Aspect Angle during the mission due to repointing to the different science targets. In the case of \theia, the telescope metrology subsystem to monitor perturbations to the telescope geometry was based on a concept of a series of simple and independent linear displacement interferometers installed between the telescope mirrors and organized in a virtual hexapod configuration.

\subsubsection{Mission configuration and profile}
\label{sec:miss-conf-prof}

The time baseline to properly investigate the science topics of this white paper would be minimally 4 years, including time devoted to orbit maintenance. A total of approximately 6 months has been estimated for the orbit transfer including the spacecraft and instrument commissioning. This estimate is made from the total of $\sim 35000$ h dedicated for the scientific program, and considering that about 15 min per slew will be dedicated to reconfiguration and station-keeping, while thermal stabilization time is in addition to the slew time.

Some instrument key features of the \theia concept are presented in Fig.~\ref{fig:mission.sc.concept}. The concept is inspired on the \euclidx\  service module with a downscaled size to minimize mass and improve mechanical properties. Similarly to \euclid and \textit{Herschel} satellites, \theia's Korsch telescope is accommodated on top of the service module in a vertical position leading to a spacecraft height of about 5m. This concept allows to optimize the payload size.

\section{World-wide context of ground-based and space science}

Observations carried out with a  mission dedicated to high precision astrometry will add significant value and will benefit from a number of other ground-based and space missions operating in the 2030s and beyond, including ESA's \textit{Athena}, \textit{PLATO}, \euclid\ and \gaia, ESO's \textit{MICADO} and \textit{Gravity}, \textit{CTA}, \textit{SKA}, \textit{JWST} and \textit{LSST}. For example:
\begin{itemize}
\item \textbf{\jwst:} Estimates suggest that \jwst will be able to detect Lyman Break galaxies with absolute magnitudes as faint as $M_{\rm{UV}} \sim -15$ at $z\sim7$, corresponding to halo masses of about $10^{9.5} \ M_{\odot}$. The combination of a high precision astrometry mission and the \jwst's observations will enable unambiguous tests of DM.
\item \textbf{\textit{PLATO}:} \plato will look at planetary transits and star oscillations in two fields (each covering 2250 deg$^2$), for 2-3 years each, in host stars brighter than 16 mag.  \textit{PLATO} high cadence continuous monitoring of its target stars will provide information on the internal structure of the stars, allowing determination of their stellar ages and masses. A high precision astrometry mission will benefit from \textit{PLATO} characterization of many of the astrometry mission's core star samples.  For close `\textit{PLATO}' stars where transits were observed this astrometry mission can measure additional inclined planets.
\item \textbf{SKA:} SKA aims to use radio signals to look for building blocks of life (e.g. amino acids) in Earth-sized planets.   A high precision astrometry will  identify target planets  from their astrometric "wobble" that can be followed-up spectroscopically with the SKA. Furthermore, SKA aims to use its immensely fast sky coverage to detect transients, such as supernovae and gamma ray bursts. With its precise astrometry, \theia will help study the specific locations of such events in stellar clusters.
\item \textbf{CTA: } The Cherenkov Telescope Array (CTA) in the Northern and Southern Hemispheres will carry out measurements of the gamma-ray flux with almost complete sky coverage and unprecedented energy and angular resolution, in the $\sim$ [0.02,100] TeV energy range. The sub-microarcsecond performance of a high precision astrometry mission allow us investigating the so-called J-factor which corresponds to the brightness of the gamma-ray flux in dSphs and thus determines the prime candidates for CTA's observations. CTA also aims at observing star forming systems over six orders of magnitude in formation rate, to measure the fraction of interacting cosmic rays as a function of the star-formation rate. By combining high precision astrometry and CTA measurements, we will better understand the relative importance of cosmic rays and DM in places where star-formation is important. Furthermore, a small number of black-hole and neutron star binary systems in our Galaxy is known to emit gamma-rays. The mechanism by which the particle acceleration is achieved is not well-understood. The sub-microarcsecond performance of a high precision astrometry mission al\-low us probing the velocity structure of the nearby gam\-ma-ray bright radio galaxies of NGC\,1275, IC\,310, M\,87 and Cen\,A, which combined with CTA's observations will enable important astrophysics breakthroughs.
\end{itemize}

\section{Technology challenges for high precision astrometry}
\label{sec:cont-other-miss}

\subsection{Spacecraft technology and cost}

There have been several propositions for a space mission dedicated to high precision astrometry: a 6 meter baseline visible interferometer on a single satellite like SIM or SIM-Lite \citep{2008SPIE.7013E..4TG}; a single mirror off-axis parabola 1 meter diameter telescope based on two spacecraft, one carrying the telescope mirror and the other the focal plane like the NEAT \citep{2012ExA....34..385M}; or a single-mirror telescope like \theia \citep{2016SPIE.9904E..2FM, Boehm2017}. The variety of the concepts shows that there are areas of progress on spacecraft technologies, especially concerning \textbf{formation flying, actively-controlled large structure interferometers}.

One interesting potential solution to be considered is the nanosat technology and the cost reduction that is linked to it. There is a huge cost difference between cubesats (< 10\,M\euro{}) and ESA M class mission (400~500\,M\euro{}) or NASA MIDEX/Discovery mission also 300~500\,M\$. The cubesat technology has matured and more than hundreds are launched every year. That technology has now crept into micro-sats that are up to 200\,kg and spacecraft bus of this category are now $<5$\,M\euro{}, while only a few years ago they were $\sim40$\,M\euro{}.   Because of their low cost and the high number of flying satellites, this technology has now demonstrated 5 year typical lifetime, comparable to a more expensive traditional spacecraft. In any case, all the price scales will change between now and the epoch when Voyage 2050 will be implemented, and that includes flying heavier payloads (SpaceX is pushing the launcher cartel prices down, for instance). 
  
\subsection{Detection}

Presently, two detector technologies are used: CCD or CMOS. CMOS detectors present a high quantum efficiency over a large visible spectral band that can also reach infrared wavelengths depending on the sensitive layer. CMOS detectors  also have programmable readout modes, faster read\-out, lower pow\-er, better radiation hardness, and the ability to put specialized processing within each pixel. On the other hand there are many known detector systematics, even for ad\-van\-ced detectors  as the Teledyne H4RG10. The main challenging effects are the following ones: fluence-dependent PSF, correlated read noise, inhomogeneity in electric field lines and persistence effects \citep[e.g.][]{Simms2009}. All mission proposals so far were based on CCD technology, but detector evolution will certainly take place on the context of any mission concept to answer the challenges being posed by the Voyage 2050 white papers.

If a \theia-like mission is selected for the 2040’s, detector technology might be different from anything we have in place nowadays. The main requirements are small pixels, low read-out noise on large format focal plane and mastering intrapixels effects in order to reach the highest precision astrometry. It should be noticed that the development of European detector technology for low-RON and large-format IR and visible detector matrices, like the Alfa detector that ESA is undertaking with Lynred, is of high interest for our science cases.
  
\subsection{Metrology}

Traditionally systematic errors have been the major challenge $\mu$as-level astrometry from space. Astrometric accuracy has a lot in common with photometric accuracy, and the technology development that proceeded the Kepler mission demonstrated $\sim 10^{-5}$ relative photometry.  Similar advances have been made in detector calibration for astrometry \citep{2016SPIE.9904E..5GC}.  Photons from stars carry the astrometric information at exquisite precision, systematic errors are imparted when those photons strike the telescope optics and also when they are detected by the focal plane array. The calibration of optical field distortion using reference stars is a technique that is perhaps a century old and used on ground and space-based telescopes.

Metrology laser feed optical fibers placed at the back of the nearest mirror to the detectors can be used to monitor distortions of the focal plane array, and to allow the associated systematic errors to be corrected \citep{2016SPIE.9904E..5GC}. Such detector calibration at $10^{-6}$pixel levels should be continued. In addition to measuring the FPA physical shape, the rest of the telescope needs monitoring to control time-variable aberrations at sub $\mu$as level. Even at very stable space environments such as L2, the telescope geometry is expected to vary for different reasons: structural lattice reorganization (as the micro-clanks observed in ESA/\gaia), outgassing and most importantly, thermo-elastic effects due to the necessary variation of the Solar Aspect Angle during the mission for pointings to the different science targets. A telescope metrology subsystem based on a concept of linear displacement interferometers installed between the telescope mirrors, with the role to monitor perturbations to the telescope geometry might be required and developed.  Existing space based interferometers from TNO, as the ESA/\gaia Basic Angle Monitor are already capable of reaching more precise measurements than those required by \theia/M5 -- BAM can perform $\sim1.5$ pm optical path difference measurements \citep{doi:10.1117/12.2026928}. A Thales telemeter developed for CNES can reach $\sim100$ pm, and the Thales interferometer produced for the MTG (Meteosat Third Generation) satellite can reach 1 nm per measurement \citep{Scheidel2011} -- higher precisions can be reached by averaging over many measurements.

For telescopes that do not have high level stability levels, there are some alternatives. One is the diffractive pupil concept that puts a precision array of dots on the primary, which produces a regular pattern of dots in the focal plane. One way to use the diffractive pupil is to look at a very bright star (0 mag) and record the diffraction pattern interspersed with observations of a much dimer target star ($\sim7$\,mag).  The diffractive pupil can also be used during science observations. But when the targetstar is $\sim7$\,mag photon noise of the diffracted light can be significantly higher than the photon noise of the reference stars ($\sim11-14$\,mag).

\section*{Conclusion}

To solve fundamental questions like
\begin{itemize}[noitemsep,label=--]
\item ``\emph{What is the nature of dark matter?}''
\item ``\emph{Are there habitable exo-Earths nearby?}''
\item ``\emph{What is the equation of state of matter in extreme environments}?''
\item ``\emph{Can we put direct constraints on cosmological models and dark energy parameters?}''
\end{itemize}
many branches of astronomy need to monitor the motion of faint objects with significantly higher precision than what is accessible today. \textbf{Through ultra-precise micro-arcsecond relative astrometry, a high precision astrometry space mission will address the large number of prime open questions that have been detailed in this white paper}.

The scientific requirements points toward a space mission that is relatively simple: a single telescope, with metrology subsystems and a camera. Such a mission can fit as a M-class mission, or even at lower level depending on the final accuracy which is aimed at.

Some technological challenges must be tackled and advanced:  the spacecraft, the focal plane detector and the metrology. We believe that these challenges can be mastered well before 2050 and that they will open the compelling scientific window of the faint objects in motion.

\clearpage
\onecolumn
\setlength{\bibsep}{0pt plus 0.3ex}
\label{sec:references}
\bibliographystyle{aa}
\bibliography{hpawp-bibliography}

\clearpage
\section*{Core proposing team}
Sorted by alphabetical order:
\begin{center}
\begin{tabular}{|l|l|c|}
  \hline
  Name &Affiliation &Country\\
  \hline
  U.~Abbas &INAF/Obs. Torino &IT\\			   
  J.~Alves &U. Vienna &AT\\
  C.~Boehm &U. Sydney &AU\\
  W.~Brown &CFA Harvard &US\\
  L. Chemin &U. Antofogasta &CL\\
  A.~Correia &U. Coimbra &PT\\
  F.~Courbin &EPFL \& Ob. Geneva &CH\\
  J.~Darling &U. Colorado &US\\
  A.~Diaferio &U. Torino/INFN &IT\\
  M.~Fortin &Copernicus Astronomical Center &PL\\
  M.~Fridlund &Leiden Obs., &NL \\
                      &\& Chalmers Univ. &SE\\
  O.~Gnedin &U. Michigan &US\\
  B.~Holl &U. Geneva &CH\\
  A.~Krone-Martins &CENTRA/U. Lisboa &PT\\
  A.~L\'eger &IAS/U. Paris Sud &FR\\
  L.~Labadie &U. Cologne &DE\\
  J.~Laskar &IMCCE/Obs. Paris &FR\\
  F.~Malbet &IPAG/U. Grenoble Alpes &FR\\
  G.~Mamon &IAP &FR\\
  B.~McArthur &U. Texas &US\\
  D.~Michalik$^*$ &ESA/ESTEC &NL\\
  A.~Moitinho &CENTRA/U. Lisboa &PT\\
  M.~Oertel &LUTH/Obs. Paris/CNRS &FR\\
  L.~Ostorero &U. Torino/INFN &IT\\
  J.~Schneider &Obs. Paris &FR\\
  P.~Scott  &Imperial College London, &UK\\
                 & \& U. Queensland, &AU\\
  M.~Shao &JPL/NASA &US\\
  A.~Sozzetti &Obs. Torino/INAF &IT\\
  J.~Tomsick &SSL Berkeley &US\\
  M.~Valluri &U. Michigan &US\\
  R.~Wyse &Johns Hopkins U. &US\\
  \hline
  \multicolumn{3}{l}{$^*$ \small \emph{ESA Research Fellow}}
\end{tabular}
\end{center}
Acknowledgments to the contributors to the \theia proposal for M5 \citep{Boehm2017}.
\end{document}